\documentclass{mn2e}

\usepackage{amsmath}
\usepackage{graphicx}
\usepackage{latexsym}

\title[Solar activity forecast with a dynamo model]{Solar activity forecast with a dynamo model}
\author[J.\ Jiang, P.\ Chatterjee and A.\ R.\ Choudhuri]{Jie Jiang$^{1}$ \thanks{E-mail:
jiangjie@ourstar.bao.ac.cn},  Piyali Chatterjee$^{2}$ and Arnab Rai Choudhuri$^{1, 2}$\\
$^{1}$National Astronomical Observatories, Chinese Academy of Sciences, Beijing 100012, China.\\
$^{2}$Department of Physics, Indian Institute of Science, Bangalore- 560012, India.}

\begin{document}
\date{}
\maketitle
\begin{abstract}
Although systematic measurements of the Sun's polar magnetic field exist only from
mid-1970s, other proxies can be used to infer the polar field at earlier times.
The observational data indicate a strong correlation between the polar field
at a sunspot minimum and the strength of the next cycle, although the strength
of the cycle is not correlated well with the polar field produced at its end.
This suggests that the Babcock--Leighton mechanism of poloidal field generation
from decaying sunspots involves randomness, whereas the other aspects of the
dynamo process must be reasonably ordered and deterministic.  Only if the magnetic
diffusivity within the convection zone is assumed to be high (of order $10^{12}$ cm$^2$
s$^{-1}$), we can explain the correlation between the polar field at a minimum and
the next cycle.  We give several independent arguments that the diffusivity must be
of this order.  In a dynamo model with diffusivity like this, the poloidal field
generated at the mid-latitudes is advected toward the poles by the meridional
circulation and simultaneously diffuses towards the tachocline, where the
toroidal field for the next cycle is produced.
To model actual solar cycles with a dynamo model having such high
diffusivity, we have to feed the observational data of the poloidal field at
the minimum into the theoretical model.  We develop a method of doing this in
a systematic way.  Our model predicts that cycle~24 will be a very weak cycle.
Hemispheric asymmetry of solar activity is also calculated with our model and
compared with observational data.

\end{abstract}


\section{Introduction}

During the last few decades, solar physicists have attempted to predict the strength
of every solar cycle a few years before its onset. When such attempts were made
to predict the last cycle~23 in the mid-1990s, there did not yet exist sufficiently
sophisticated and detailed models of the solar dynamo.  So most of the attempts were
primarily based on various precursors which were
expected to give an indication of the next solar cycle.
Solar dynamo theory has progressed enormously in the last few years and first
attempts are now made to predict the next cycle~24 from dynamo models.
Dikpati et al.\ (2006) and
Dikpati \& Gilman (2006) have predicted that the cycle~24 will be the strongest cycle
in 50 years.  On the other hand, Choudhuri et al.\ (2007) have used a different model
and different methodology to conclude that the cycle~24 will be the weakest in 100
years.  Irrespective of which prediction turns out to be correct, the next cycle~24
should be regarded as a historically important cycle in the evolution of solar
dynamo theory---as the first cycle for which detailed dynamo predictions could be
made.  In view of these contradictory predictions, it is clear that cycle prediction
is a fairly model-dependent affair.  Since there are still many uncertainties in
solar dynamo models (Choudhuri 2007a), it may be worthwhile to analyze the physical
basis of the solar cycle prediction carefully, rather than having too much faith on
predictions from any particular model.

Amongst the so-called precursor methods, the most popular method first proposed by
Schatten et al.\ (1978) is to use the polar magnetic
field at the preceding minimum as a the
precursor for the next maximum.  Since the polar field is weak at the present
time, Svalgaard et al.\ (2005) and Schatten (2005) have predicted a weak cycle~24.
This prediction is in agreement with the
dynamo-based prediction of Choudhuri et al.\ (2007),
but not with the prediction of Dikpati et al.\ (2006). One important question
before us is whether this polar field method is reliable.  The next important
question is whether dynamo models provide any support for this method.

Since systematic polar field measurements are available only from mid-1970s, we
so far have only 3 data points indicating a strong correlation between the polar field
at a minimum and the next maximum, as we discuss in the \S2.  Even if the correlation
may appear good, it can be argued that 3 data points do not constitute a good
statistics.  However, we can use various other proxies like polar faculae
(Sheeley 1991) and positions of filaments (Makarov et al.\ 2001) to infer
polar fields from the beginning of the 20th century, and we argue in \S2 that
there are good reasons to have faith on the polar field method for predicting
solar cycles.

We now come to the question whether the polar field method can be justified
from dynamo theory.  Both the predictions of Dikpati et al.\ (2006) and Choudhuri
et al.\ (2007; hereafter CCJ) are based
on flux transport dynamo models.  In these models, the
main source of poloidal field is the Babcock--Leighton process in which tilted
bipolar regions on the solar surface give rise to a poloidal field after their
decay.  This poloidal field is advected by the meridional circulation towards
the pole, to create a strong (i.e. of order 10 G) polar field at the time of the
solar minimum.  The conventional wisdom is that this polar field is then
advected downward by the meridional circulation to bring it to the tachocline,
where it is stretched by the differential rotation to create the toroidal
field, ultimately leading to active regions due to magnetic buoyancy. The
correlation between the polar field at the minimum and the strength of the next
maximum can be explained if this polar field can be brought to the
mid-latitude tachocline by the time of the next maximum.  Since a maximum
comes about 5 years after a minimum, the advection time from the pole to the
mid-latitude tachocline has to be of order 5 years if this explanation is to
work.  Although we do not have any direct observational data on the nature
of the meridional circulation in the lower half of the convection zone, the
time scale of this circulation seems to set the period of the dynamo
(Dikpati \& Charbonneau 1999; Hathaway et al.\ 2003)
and we cannot vary the amplitude of the
meridional circulation at the bottom of the convection zone too much if we wish
to reproduce various observed features of the solar cycle, especially
its period.  Charbonneau
\& Dikpati (2000) pointed out that the advection time in their model was of
order 20 years and led to a correlation of the polar field at the end of
cycle {\it n} with the strengths of the cycles $n+2$
and $ n+3$ rather than the cycle $n+1$.
The same is presumably true for the simulations of Dikpati et al.\ (2006)
and Dikpati \& Gilman (2006).

In the CCJ model, a sudden change in the polar field at the time of a minimum
has a prominent effect on the next maximum coming only after 5 years, as can
be seen in Fig.\ 2 of CCJ.  In the caption of Fig.\ 2, CCJ offered an explanation
by suggesting that the advection time in their model was shorter than that
in the models of Dikpati and co-workers.  We now feel that this explanation
was erroneous.  When we find two phenomena {\it A} and {\it B} correlated,
our first guess usually is that the earlier phenomenon is the cause of the
later phenomenon.  But an alternative explanation is also possible.  If both
{\it A} and {\it B} are caused by {\it C} which took place earlier than both
{\it A} and {\it B}, then also it is possible for {\it A} and {\it B} to
appear correlated. We now believe that the polar field at the minimum and
the strength of the next maximum are
correlated not because the polar field was the direct
cause of the next maximum by being advected from the pole to the tachocline.
Rather, they appear correlated because both of them arise from the poloidal
field produced by the Babcock--Leighton process in the mid-latitudes.
Let us explain this point with Fig.~1.

\begin{figure}
\centering{\includegraphics[width=6cm]{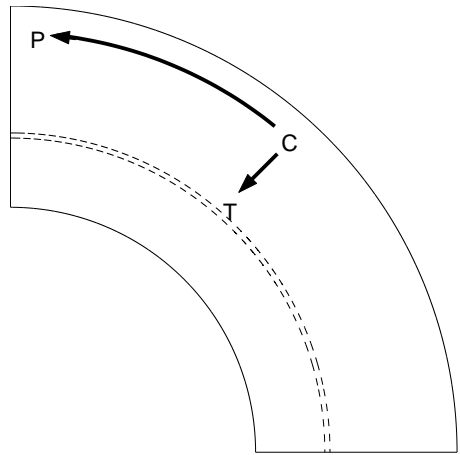}}
\caption{A sketch
indicating how the poloidal field produced at {\it C} during a
maximum gives rise to the polar field at {\it P} during the
following minimum and the toroidal field at {\it T} during the next
maximum.}
\end{figure}

During a maximum, the poloidal field is created by the Babcock--Leighton
process primarily in the region {\it C} indicated in Fig.~1. This field is
advected by the meridional circulation to the polar region {\it P} to produce
the polar field at the minimum.  If diffusion is important, then the poloidal
field produced at {\it C} also keeps diffusing.  The diffusion in our dynamo
model is stronger than in the model of Dikpati and co-workers, as pointed
out by Chatterjee et al.\ (2004) and Chatterjee \& Choudhuri (2006). So,
in a few years, the poloidal field diffuses from {\it C} to reach
the tachocline at {\it T} in our model, which will not happen in the model
of Dikpati \& Gilman (2006) where the poloidal field will be swept away from
{\it C} to {\it P} completely by the meridional circulation before it has
any chance to reach the tachocline due to the low diffusivity of that model.
Thus, in our model, the poloidal field at {\it C} produced during a maximum
gives rise to the polar field at {\it P} during the next minimum and also
the toroidal field at {\it T} which is the cause of active regions during the
subsequent maximum.  The polar field at the minimum and the strength
of the next maximum appear correlated not because one is the cause of
the other, but because both of them have the poloidal field of the previous
cycle as their cause.  If the poloidal field produced in the previous cycle
was strong, then both of these will be strong, and vice versa.

This brings us to the crucial question as to what determines the strength
of the poloidal field produced by the Babcock--Leighton process.  CCJ argued
that this process involves some randomness and the actual poloidal field produced
in the Sun at the end of a cycle will in general be different from the polar field
produced in a theoretical mean field dynamo model. The polar field of the
theoretical model at the solar minimum will be characteristic of a typical
`average' cycle. To model actual cycles, CCJ proposed that the theoretical
model should be `corrected' by feeding information about the observed poloidal
field at the minimum into the theoretical model in some suitable fashion.
CCJ had done this by using the values of DM (Dipole Moment, which is a good
measure of the polar field) at the minima computed by Svalgaard et al.\ (2005).
Since values of DM are available only from mid-1970s, this method could be
applied to model only the last few solar cycles.
While this method yielded results agreeing reasonably well for cycles~21--23,
using only one number like the DM value to characterize the poloidal field
at a minimum may seem like a drastic simplification.  Especially, if the poloidal
field produced all over the surface diffuses through the convection zone to
reach the tachocline, then we ought to feed information about poloidal field
at all latitudes into our dynamo model rather than feeding only the DM value.
One of the aims of this paper is to develop a formalism to do this.

We first discuss in \S2 whether the polar fields at minima seem
sufficiently well correlated with the strengths of the next maxima
on the basis of the observational data.  Then a brief
description of our dynamo model is provided in \S3. Then in \S4
we present some calculations done by introducing stochastic
fluctuations at the minima in dynamo models with high and low
diffusivities to show that a high-diffusivity model provides a
better fit with observational data.  Several independent arguments
in favour of a high magnetic diffusivity are put together in \S5.
Then in \S6 we discuss our methodology of processing the
observational data from Wilcox Solar
Observatory (WSO) and feeding them into the theoretical
dynamo model. Our results based on calculations with more detailed
data of poloidal field are presented in \S7. We close in \S8 with concluding
remarks.

\section{The implications of observational data}

Let us first discuss whether observational data provide good support
to the hypothesis that the polar field at the preceding minimum is a
reliable precursor for the strength of the next maximum.  Table~1 lists
the strengths of the last few cycles and the DM values at the minima at the
ends of these cycles (only the cycles for which DM values are
available are included).  The DM values are taken from Svalgaard et al.\
(2005) as discussed by CCJ. Fig.~1 of Choudhuri (2007b) plotted the
strengths of cycles~$n+1$ against the DM values at the ends of cycles~$n$.
Although we have only 3 data points, they lie very close to a straight line
implying a strong correlation.  These data points are shown by the
solid circles in Fig.~2 of this paper.
The polar magnetic field of the Sun, which was first detected by
Babcock \& Babcock (1955) and has been measured occasionally from
that time, has been systematically recorded by Wilcox Solar
Observatory (WSO) and Mount Wilson Observatory (MWO) from
mid-1970s. While we do not have systematic direct
measurements of the polar magnetic field at earlier times, the important
question is whether we can indirectly infer the values of this field at
earlier times and check if the correlation seen in the last 3 cycles
also existed in earlier cycles.

\begin{table}
\begin{tabular}{ccc}
\hline
Cycle & Maximum strength & DM value at \\
number & of the cycle $R_{\rm max}$ & the end of cycle \\
\hline
20 &110.6 & 250\\
21 &164.5 & 245.1\\
22 &158.5 & 200.8\\
23 &120.8 & 119.3 \\
\hline
\end{tabular}
\caption{Maximum strength of the cycle and the DM value at its end are
listed against cycle number.  The data for this table are taken from
Svalgaard, Cliver \& Kamide (2005).}
\end{table}

We are aware of two works which lead to the possibility of inferring
polar fields at earlier times.  Sheeley (1991) has compiled the numbers
of polar faculae seen in different years
in the white-light plates of MWO
during the period 1906--1990 and has argued
that the polar field strength has a good correlation with
the number of polar faculae.   The second work which
throws light on the evolution of the polar field during 1915--1999
is the work by Makarov et al.\ (2001),
who take the dark filaments seen in white-light plates as indicators
of neutral lines where the diffuse magnetic field on the solar surface
changes sign.  Assuming the positive and negative radial magnetic fields on
the solar surface to have values +1 G and -1 G respectively, they compute
a quantity $A(t)$ which is a measure of the Sun's large-scale magnetic
field.  Since it is the polar field which is the primary source of the
Sun's large-scale magnetic field during the solar minima, values of
$A(t)$ during the minima should be an indicator of the polar field.
Both the polar faculae number plotted in Fig.~1 of Sheeley (1991) and
$A(t)$ plotted in Fig.~1 of Makarov et el.\ (2001) have peaks at the
solar minima.  These peak values of either faculae number or $A(t)$ at
the successive minima can be taken as measures of the polar field
at these minima.  Unfortunately the polar fields estimated from the polar
faculae number and from $A(t)$ do not always agree with each other.
So one has to be very cautious in using polar field values inferred
from either of these data.

\begin{figure}
\begin{center}
\includegraphics[width=6cm]{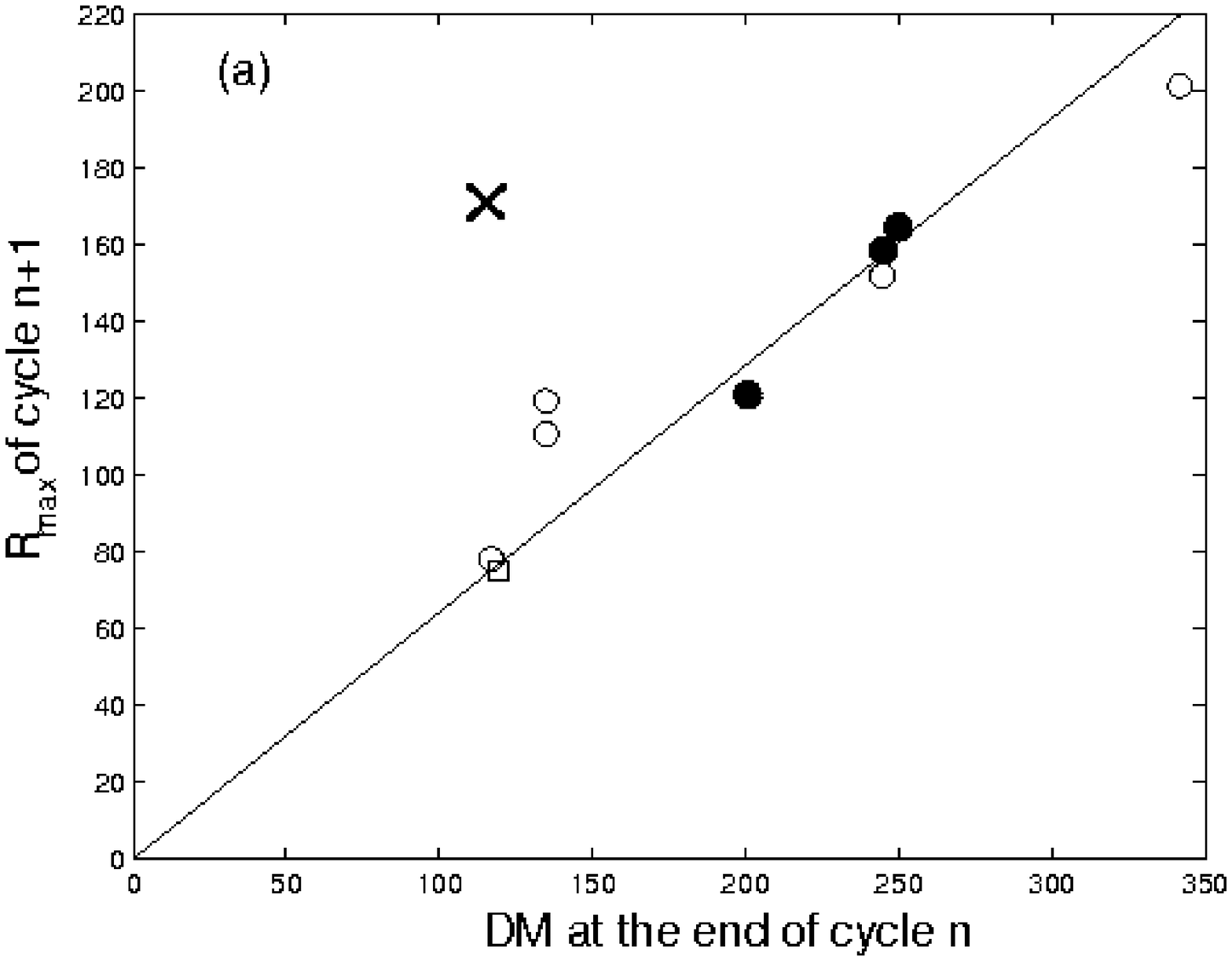}
\includegraphics[width=6cm]{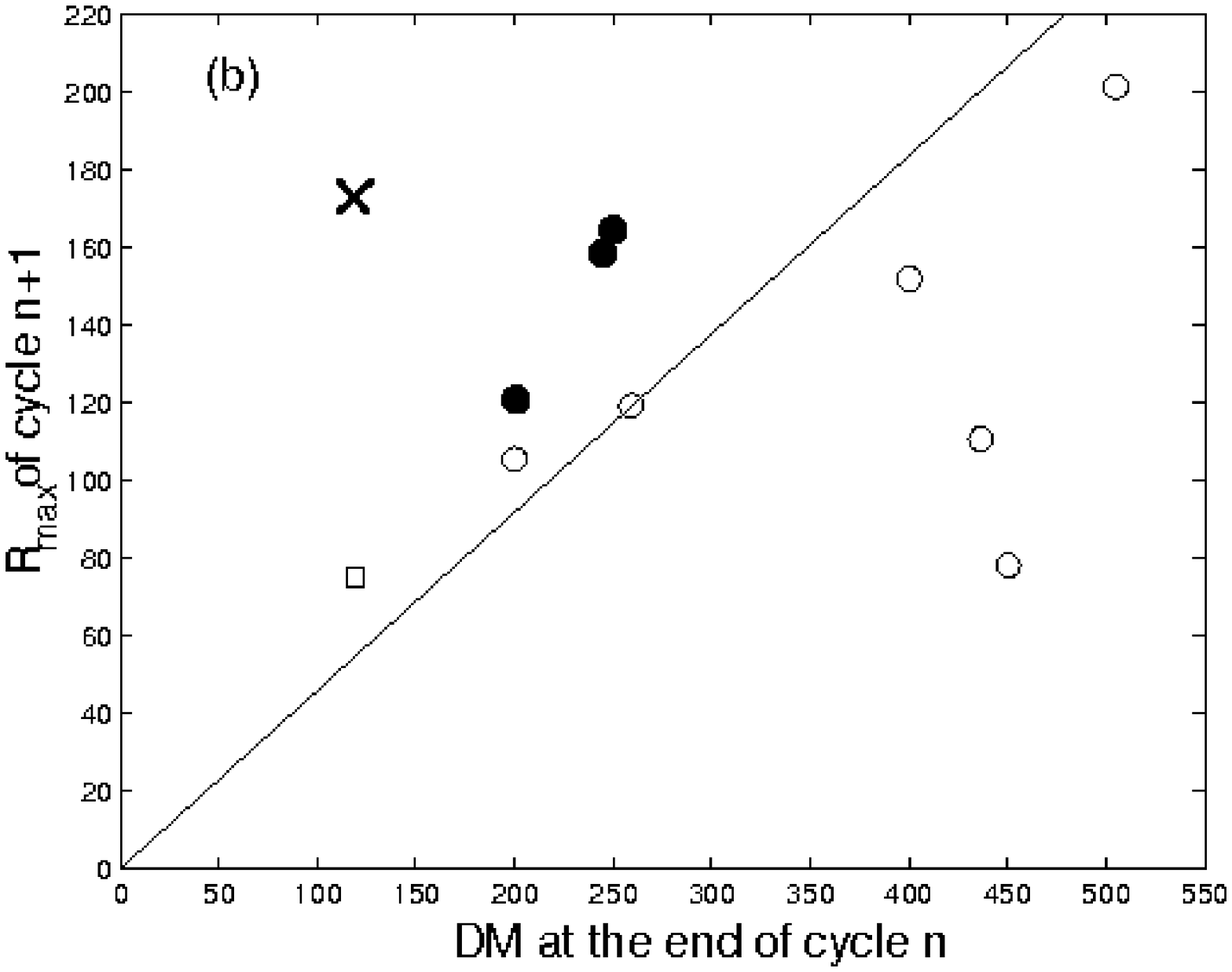}
\caption{Strengths of solar cycles plotted against the DM values of
polar fields at the preceding minima. The solid circles are based on
actual polar field measurements, whereas the open circles are based
on polar fields inferred (a) from values of $A(t)$ at the minima as
given by Makarov et al.\ (2001), and (b) from the numbers of polar
faculae at the minima as given by Sheeley (1991). The points marked
by `$\times$' and `$\Box$' indicate the data points for cycle~24
according to the models of Dikpati \& Gilman (2006) and CCJ.}
\end{center}
\end{figure}

For the minima at the ends of cycles 20, 21 and 22, we have DM values
available as well as values of $A(t)$
computed by Makarov et al.\ (2001).  If we divide DM by the peak values
of $A(t)$ at these minima, we get 19.2, 15.8 and 19.1 respectively.
Since the average of these is 18.0, we assume that we can multiply
peak values of $A(t)$ at the earlier minima to get the DM values (in $\mu$T)
at these minima.  We estimate DM values at the ends of cycles 15--19
in this way. Strengths of cycles~$n+1$ plotted against these DM values
are shown by open circles in Fig.~2(a). The
solid circles are based on the actual magnetic field measurements
at the ends of cycles 20--22. It seems that there is a
very good correlation between the polar field at the end of a cycle
and the strength of the following cycle. The points marked by `$\times$'
and `$\Box$' indicate the data points
for cycle~24 according to the models of Dikpati \& Gilman (2006) and CCJ.
Dikpati \& Gilman (2006) predict that cycle 24 will be
30-50\% stronger than cycle 23. Since cycle 23 had
$R_{\rm max} = 120.8$, this gives a range 157-181 for the
$R_{\rm max}$ value of cycle 24, with an average of 169.
We have used this value in Fig.~2(a).

If we use the peak polar faculae numbers as given in Fig.~1 of Sheeley (1991)
to estimate the polar fields at the minima, then the correlation turns
out to be considerably worse.  This is shown in Fig.~2(b).  The open circles
in this figure are obtained by assuming that the DM values (in $\mu$T) at the
minima are given by multiplying the total number of polar faculae (i.e.\
the sum of north and south polar faculae) by a factor 4.55. This is
done for the minima at the ends of cycles 14--19, whereas the solid
circles are based on actual polar field measurements at the ends of
cycles 20--22.  The correlation would have looked considerably better
if two data points at the bottom right did not exist.  These two data
points correspond to the minima around 1923 and 1964. These minima were
followed by the two weakest cycles in the past century.  According
to Fig.~1 of Sheeley (1991), the polar faculae counts during these
minima were reasonably high, suggesting that the polar fields would
have been strong and thus offsetting the two data points in Fig.~2(b).  On
the other hand, values of $A(t)$ at these minima, as shown in Fig.~1
of Makarov et al.\ (2001), were quite low.  This suggests weaker magnetic
fields at these minima, which would bring the two data points much closer
to the correlation line.  We shall probably never know for sure whether the
polar fields at these minima were actually weak or
strong.  This shows that using
other parameters as proxies of the polar field can be problematic.
Only when we have reliable polar magnetic measurements for several cycles,
we shall be able to determine really how good a correlation exists between
the polar fields at the minima and the strengths of the next cycles.
Errors in the polar field estimate probably make the correlation look
worse than what it actually is.  For example, if we had actual polar
field measurements for the two data points at the bottom right of
Fig.~2(b), probably these points would lie closer to the correlation
line.  In spite of various uncertainties,
Figs.~2(a) and 2(b) suggest that the correlation between the
polar field at a minimum and the strength of the next maximum
is reasonably good.  It may be noted that Makarov et al.\ (1989)
found a correlation between the polar faculae number and the
sunspot number about 6 years later.

\begin{figure}
\centering{\includegraphics[width=6cm]{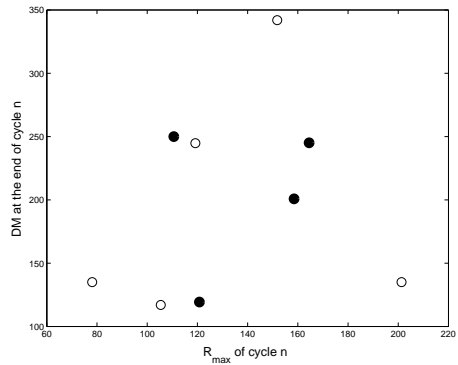}}
\caption{DM values of polar fields at the minima plotted against the
strengths of the previous solar cycles. The solid circles are based
on actual polar field measurements, whereas the open circles are
based on polar fields inferred from values of $A(t)$ at the minima
as given by Makarov et al.\ (2001).}
\end{figure}

Fig. 3 plots the DM values of the polar field at the ends of cycles
against the strengths of those cycles.  Again, the 4 solid circles are based
on actual polar field measurements, whereas the 5 open circles are
based on DM values inferred from $A(t)$ peak values.  Neither the solid
circles, nor the open circles show much correlation.  It is clear that
the strength of a cycle does not determine the polar field produced
at the end of the cycle, implying that the generation of the poloidal
field involves randomness. The lack of correlation in Fig.~3 can be
taken as a justification behind the assumption of CCJ that the polar
field at the end of a cycle cannot be inferred from the sunspot data
of the cycle and has to be fed into the theoretical model by using
actual observational data.  On the other hand, Dikpati \& Gilman (2006)
have used the sunspot area data as the
completely deterministic source of poloidal field in
their model.  According to our judgment, it is wrong to assume
a process which clearly involves randomness and is poorly correlated
to be deterministic.  Using suitably averaged sunspot area data,
Cameron \& Sch\"ussler (2007) found a rather intriguing correlation
between the theoretically computed magnetic flux crossing the equator
at the minimum and the strength of the next maximum, for the last
few cycles.  They suggested this as a possible reason how Dikpati \&
Gilman (2006) ``predicted'' the past cycles.  However, this correlation
virtually disappeared when Cameron \& Sch\"ussler (2007) used more detailed
sunspot data rather than the smoothed data.

\section{The standard dynamo model}

The toroidal field of the dynamo is universally believed to be
produced in the tachocline.  A very influential idea for the generation
of the poloidal field is the $\alpha$-effect, which assumes that
the toroidal field is twisted by helical turbulence to give rise
to the poloidal field (Parker 1955; Steenbeck et al.\ 1966). When
flux tube rise simulations established that the toroidal field at
the bottom of the solar convection
zone (hereafter SCZ) has to be much stronger than the equipartition
field (Choudhuri \& Gilman 1987; Choudhuri 1989; D'Silva \& Choudhuri
1993; Fan et al.\ 1993), it became clear that the traditional
$\alpha$-effect will be suppressed.  This has led several dynamo
theorists in recent years to invoke an alternative idea of
poloidal field generation from the decay of tilted active regions
proposed by Babcock (1961) and Leighton (1969). The meridional
circulation, which is poleward near the surface and equatorward
at the bottom of SCZ, has to play an important role in such
dynamo models called `flux transport dynamos' (Wang et al.\ 1991).
Two-dimensional flux transport dynamo models were first constructed
by Choudhuri et al.\ (1995) and Durney (1995).

Most of our calculations are based on the dynamo model presented
by Nandy \& Choudhuri (2002) and Chatterjee et al.\ (2004).  The
readers are advised to consult either Chatterjee et al.\ (2004)
or Choudhuri (2005) for the full details of the model.  Here we
present only the salient features.
The basic equations for the standard axisymmetric $\alpha\Omega$
solar dynamo model are
\begin{equation}
\frac{\partial A}{\partial
t}+\frac{1}{r\sin\theta}(\textbf{v}\cdot\nabla)(r\sin\theta
A)=\eta_p(\nabla^2-\frac{1}{r^2\sin^{2}\theta})A+\alpha B,
\end{equation}
\begin{eqnarray}
\frac{\partial B}{\partial t}+\frac{1}{r}[\frac{\partial}{\partial
r}(rv_rB)+\frac{\partial}{\partial\theta}(v_\theta B)]
=\eta_t(\nabla^2-\frac{1}{r^2\sin^{2}\theta})B\nonumber\\
+r\sin\theta(\textbf{B}_p\cdot\nabla)\Omega+\frac{1}{r}\frac{d\eta_t}{dr}\frac{\partial}{\partial
r}(rB),
\end{eqnarray}
where $B(r,\theta,t)\bf{e}_\phi$ and
$\nabla\times[A(r,\theta,t)\bf{e}_\phi]$ respectively correspond to
the toroidal and poloidal components. Here $\textbf{v}$ is the
meridional flow, $\Omega$ is the internal angular velocity and
$\alpha$ describes the generation of poloidal field from the
toroidal field.  The turbulent diffusivities
for the poloidal and toroidal field are denoted by $\eta_p$ and $\eta_t$.
Since turbulence has less effect on the stronger toroidal field,
we in principle allow $\eta_p$ and $\eta_t$ to be different.
Magnetic buoyancy is treated by removing a part of $B$ from
the bottom of SCZ to the top whenever $B$ exceeds a critical
value, as discussed by Chatterjee et al.\ (2004, \S2.6).
Such a treatment of magnetic buoyancy coupled with $\alpha$
concentrated at the top of SCZ captures the essence of the
Babcock--Leighton process of poloidal field generation
(Nandy \& Choudhuri 2001). Although the period of a flux
transport dynamo is determined mainly by the
meridional flow speed (Dikpati \& Charbonneau 1999; Hathaway
et al.\ 2003) and this flow speed is known to vary,
the detailed time variation is known
only since 1996 (Gizon 2004). Hence, we adopt a steady meridional
flow speed. Chatterjee et al.\ (2004, \S2) and Choudhuri (2005)
describe how the various parameters
$\bf{v}$, $\Omega, \eta_p, \eta_t$ and $\alpha$ were specified
to produce what they called their {\em standard} model, of
which the solution was presented in \S4 of
Chatterjee et al.\ (2004). The period of this
standard model was about 14 years. We change $\bf{v}$ along
with some other parameters to get a period of 10.8 years. The old
values and the changed values of the parameters are listed in Table 1.
The dynamo model with the changed values giving a period of
10.8 years is referred to our {\em standard1} model.  Most of
our high-diffusivity calculations in this paper are done with
this {\em Standard1} model.

Since the diffusion of the poloidal field is going to play a
very important role, we write down the expression of $\eta_p$
which we use, although the reader is referred to Chatterjee
et al.\ (2004) or Choudhuri (2005) for the expressions of the
other parameters.  We take $\eta_p$ to be given by
$$\eta_{p}(r) = \eta_{RZ} + \frac{\eta_{SCZ}}{2}\left[1 + \mbox{erf} \left(\frac{r - r_{BCZ}}
{d_t}\right) \right]. \eqno(3)$$
Here $\eta_{SCZ}$ is the turbulent diffusivity inside the convection
zone, which is taken as $2.4 \times 10^{12}$ cm$^2$ s$^{-1}$ in the
{\em Standard1} model.  The diffusivity $\eta_{RZ}$ below the bottom of
SCZ is assumed to have a rather value of $2.2 \times 10^{8}$ cm$^2$ s$^{-1}$.
A plot of $\eta_p$ can be seen in Fig.~4 of Chatterjee et al.\ (2004).

\begin{table}
\caption[]{The original values of the parameters in the standard
model (\S4 of Chatterjee et al.\ 2004) along with the changed values we
use now. The first four parameters control the amplitude,
penetration depth, equatorial return flow thickness and the
position of the inversion layer of the meridional circulation,
respectively. The half width of tachocline is denoted by
$d_{tac}$.}
  \begin{center}\begin{tabular}{c|c|c}
 \hline
Parameter & Standard Model & This Model \\
\hline\hline
$v_0$ & $-29$ m s$^{-1}$ &  $-35$ m s$^{-1}$ \\
$R_p$ & $0.61 R_{\odot}$ & $0.63 R_{\odot}$\\
$\alpha_0$ & 25 ms$^{-1}$ & 22.5 ms$^{-1}$ \\
$\beta_2$ & $1.8\times10^{-8}$m$^{-1}$ & $1.3\times10^{-8}$ m$^{-1}$ \\
$r_0$ & $0.1125 R_{\odot}$ & $0.1184 R_{\odot}$ \\
$d_{tac}$ & $0.025R_{\odot}$ & $0.015R_{\odot}$\\
\hline
\end{tabular}
  \end{center}
\end{table}

We carry on our calculations with the solar dynamo code {\em Surya} developed
at the Indian Institute of Science.  This code and a detailed guide (Choudhuri
2005) for using it are made available to anybody who sends a request to
Arnab Choudhuri (e-mail address: arnab@physics.iisc.ernet.in). This code
has not only been used for doing several dynamo calculations (Nandy \&
Choudhuri 2002; Chatterjee et al.\ 2004; Choudhuri et al.\ 2004; Chatterjee
\& Choudhuri 2006; CCJ), a modified version of the code has also been used to
study the magnetic field evolution in neutron stars (Choudhuri \& Konar
2002; Konar \& Choudhuri 2004).

A flux transport dynamo combines three basic process:
(i) the strong toroidal field $B$ is produced by
the stretching of the poloidal field by differential rotation
$\Omega$ in the tachocline; (ii) the toroidal field
generated in the tachocline rises due to magnetic buoyancy
to produce sunspots (active regions) and
the decay of tilted bipolar sunspots
produces the poloidal field $A$ by the Babcock--Leighton mechanism; (iii) the
meridional circulation $\bf{v}$ advects the poloidal field first to
high latitudes and then down to the tachocline at the base of the
convection zone, although we are now suggesting that diffusivity
may play a more significant role than meridional circulation in
bringing the poloidal field to the tachocline in high-diffusivity
models. CCJ argued that the
processes (i) and (iii) are reasonably ordered and
deterministic, whereas the process
(iii) involves an element of randomness, which presumably is
the primary cause of solar cycle fluctuations. Firstly, although
active regions appear in a latitude belt at a certain phase of the
solar cycle, where exactly within this belt the active regions
appear seems random. Secondly, there is considerable scatter in the
tilts of bipolar active regions  around the average
given by Joy's law (Wang \& Sheeley 1989). The action of the
Coriolis force on the rising
flux tubes gives rise to Joy's law (D'Silva \& Choudhuri 1993), whereas
convective buffeting of the flux tubes in the upper layers of the
convection zone causes the scatter of the tilt angles
(Longcope \& Fisher 1996; Longcope \& Choudhuri 2002).
Since the poloidal field generated from an active region by the
Babcock--Leighton
process depends on the tilt, the scatter in the tilts introduces a
randomness in the poloidal field generation process.
We suggest that the poloidal field at the solar minimum produced in
a mean field dynamo model is some kind of `average' poloidal field
during a typical solar minimum. The poloidal field during a
particular solar minimum may be stronger or weaker than this average
field. By feeding the observational data into the model, we have to
correct the `average' poloidal field to make the prediction of
the next cycle.

\def\Rs{R_{\odot}}

CCJ corrected the average poloidal field at a minimum by the very simple
method of changing the values of $A$ above $0.8 \Rs$ in accordance
with the DM value at that minimum.  The advantage of this method
is that it is extremely easy to implement.  It is, however, a
drastic over-simplification to represent the poloidal field at the
minimum by a single number.  Especially, in the high-diffusivity
models, if the poloidal field diffuses
downward at different latitudes instead of being advected
to the pole first, then it may be important to develop a
methodology of feeding values of poloidal field at different
latitudes into the theoretical model instead of using only
the value of DM.
Wilcox Solar Observatory (WSO) has regularly measured the
line-of-sight component of the magnetic field using the $5250~ \rm
{\AA}~Fe~ I$ line since the later part
of 1976. This component can be taken as a simple
projection of the radial field. We will discuss in \S6 how to analysis WSO
data and connect them with our dynamo model.  We would like to point
out that the quality of the data should be sufficiently high to ensure
that our methodology gives meaningful results.  In \S7 we shall
present our results obtained with WSO data.  When we tried to use
the data from National Solar Observatory
(NSO), we were completely unable to model the past cycles properly.
Before we present the results obtained with detailed poloidal field
data, we use the simpler method of CCJ for updating the poloidal
field in the next section to highlight the differences between
dynamo models with high and low diffusivities.

Even before the development of realistic flux transport dynamo models,
Choudhuri (1992) suggested that the stochastic fluctuations
around the mean values of various quantities may be the source of
irregularities in the solar cycle.  This idea was further explored
by several other authors (Moss et al.\ 1992; Hoyng 1993; Ossendrijver et al.
1996; Mininni \& Gomez 2002).
Now we identify the randomness in the Babcock--Leighton process as
the source of stochastic fluctuations in the solar dynamo.

\section{Contrasting behaviours of dynamos with high and low diffusivities}

In \S2 we have seen that there is reasonably convincing
observational evidence that the
strength of the polar field at a minimum plays an important role in
determining the strength of the next maximum coming about 5 years later.
The results of CCJ reproduce this observed pattern.  Fig.~1 of the
present paper and the
accompanying text explains how this happens. It is the poloidal field
cumulatively generated during the declining phase of the cycle which
is responsible for both the polar field at the end of the cycle
(produced by poleward advection of this poloidal field from mid-latitudes)
and the maximum of the next cycle (since the toroidal field is generated
from this poloidal field which has diffused to the bottom of the convection
zone).   To support our assertion, the left column of Fig.~4 shows how
the poloidal field lines evolve in our {\em Standard1} model with diffusivity
on the higher side.  For the sake
of comparison, the right side of Fig.~4 shows the poloidal field evolution
in a low-diffusivity model, which we shall discuss later.
Since we are not concerned with parity issue here, most of the calculations
in this Section are done in a hemisphere with boundary conditions at
the equators appropriate for a dipolar solution.

\begin{figure}
\centering{\includegraphics[width=7cm]{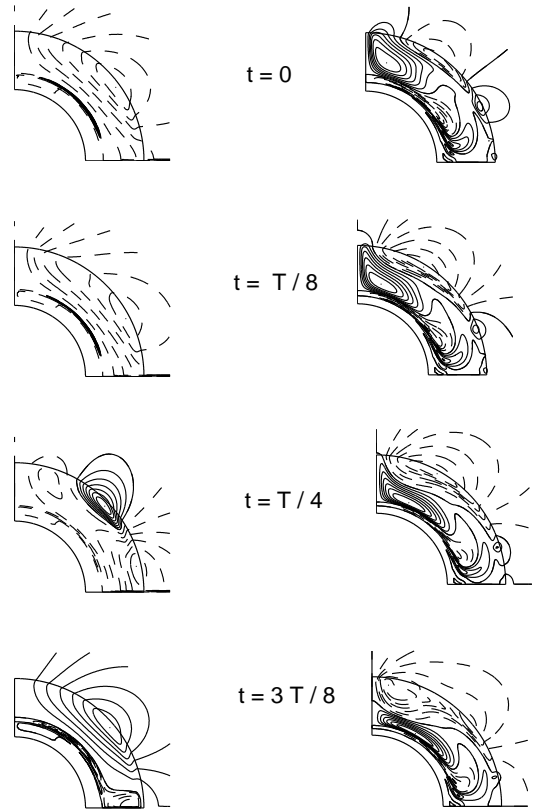}}
\caption{Evolution
of the poloidal field.  The left column corresponds to our
high-diffusivity {\em Standard1} model.  The right column
corresponds to the low-diffusivity of model of Dikpati \&
Charbonneau (1999), except that we have taken $u_0 = 20$ m
s$^{-1}$.}
\end{figure}

We have seen in Fig.~2 of CCJ that, if the poloidal field above $0.8\Rs$
is suddenly changed at a minimum, then the maximum coming soon after that
and the subsequent maximum are both affected.  From this, we expect that
the strength of a maximum should depend on the polar field strengths at the
two preceding minima.  Thus, while the polar field at the immediately
preceding minimum should not determine the strength of the maximum
completely, we expect our theoretical model to show a good correlation
between the polar field of a minimum and the strength of the next maximum,
as we see in the observational data discussed in \S2.  Since the poloidal
field generated at the surface has to diffuse to the tachocline in a few
years to produce this correlation, we expect that the correlation will
get worse if the diffusivity is reduced.  We carry out some numerical
experiments to test this.

\begin{figure}
\centering{\includegraphics[width=7cm]{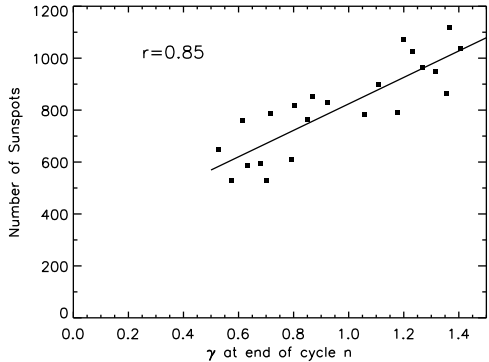}} \caption{The
strength of the maximum of cycle $n+1$ plotted against the randomly
chosen value of $\gamma$ at the end of cycle $n$. $Standard1$ model
is used to generate this plot.}
\end{figure}

\begin{figure}
\centering{\includegraphics[width=7cm]{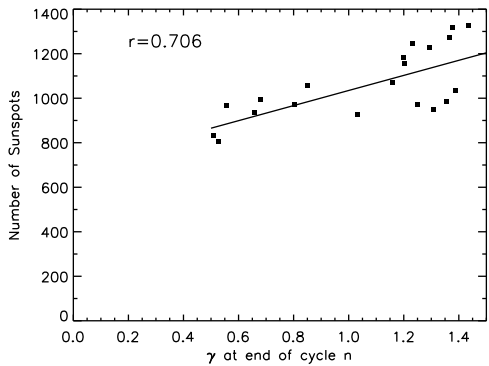}} \caption{The
strength of the maximum of cycle $n+1$ plotted against the randomly
chosen value of $\gamma$ at the end of cycle $n$. The diffusivity
$\eta_P$ is made half its value used in the $Standard1$ model to
generate this plot.}
\end{figure}

We run our dynamo code for several cycles, stopping it at every
minimum and changing the value of $A$ above $0.8\Rs$ in the
following fashion. We use a random number generating programme to
generate random numbers between 0.5 and 1.5.  We take one of these
random numbers as the factor $\gamma$ for a minimum and multiply $A$
above $0.8\Rs$ by a constant number such that the amplitude of the
poloidal field becomes $\gamma$ times the amplitude of the poloidal
field produced in an average cycle (i.e.\ when the code is run
without any interruptions).  Fig.~5 plots the strengths of the next
maxima against values of $\gamma$ chosen in  the preceding minima
(which can be regarded as indicative of the strengths of the polar
field at the minima).  We see a very good correlation as we see in
the observational data of Fig.~2.  Then we repeat this numerical
experiment by reducing the diffusivity of the poloidal field within
the convection zone to half its value, i.e. $\eta_{SCZ}$ is changed
from the value $2.4 \times 10^{12}$ cm$^2$ s$^{-1}$ used in our {\em
Standard1} model to the value $1.2 \times 10^{12}$ cm$^2$ s$^{-1}$.
All the other parameters are kept the same, except we change
$\alpha_0$ to 9 m s$^{-1}$ to make sure that the solutions remain
oscillatory.  The run with this reduced diffusivity gives Fig.~6,
where we find the correlation to be worse than what it is in Fig.~5.
It is thus clear that reducing the diffusivity in the theoretical
model leads to worsening of the correlation between the polar field
at the minimum and the strength of the next maximum.  However,
Fig.~6 still represents a case in which the poloidal field reaches
the tachocline by diffusing from the surface in a few years. If we
really want to make diffusion inefficient such that the poloidal
field cannot reach the tachocline by diffusion and has to be
advected there by the meridional circulation, then we have to reduce
the diffusivity of the poloidal component in our model by at least
one order of magnitude.  As it happens, our model does not give
oscillatory solutions if the diffusivity of the poloidal field in
the convection zone is reduced by a factor of 10 while keeping the
other parameters unchanged.  We, therefore, carry on some tests on
the model of Dikpati \& Charbonneau (1999) to study the behavior of
a dynamo with low diffusivity.

Dikpati \& Charbonneau (1999) present what they call a `reference solution'
in \S3 of their paper. The turbulent diffusivity within the convection
zone in this solution is taken to be $5 \times 10^{10}$ cm$^2$ s$^{-1}$,
which is about 50 times smaller than the value we use in our {\em Standard1}
model. We try to reproduce this reference solution with our
dynamo code {\em Surya} by changing the various parameters to what
are given in the paper of Dikpati \& Charbonneau (1999).  Especially,
we change the form of the meridional circulation to the from proposed
by van Ballegooijen \& Choudhuri (1988), which has been used by
Dikpati \& Charbonneau (1999). We also treat the magnetic buoyancy
in the non-local way as they have done.  We found that our solution
had a longer period and looked somewhat different from the solution
presented in Fig.~3 of Dikpati \& Charbonneau (1999).  However, when
we take the amplitude of the meridional circulation to be $u_0 = 20$
m s$^{-1}$ rather than $u_0 = 10$ m s$^{-1}$ as quoted by Dikpati \&
Charbonneau (1999), we get the theoretical
butterfly diagram shown in Fig.~7, which very closely
resembles  Fig.~3 of Dikpati \& Charbonneau (1999). The
evolution of the poloidal field in this model has been shown on the right
side of Fig.~4.  This can be compared with the plots given on the right
side of Fig.~2 of Dikpati \& Charbonneau (1999).  Again we find that our
plots look very similar to theirs.  It thus appears that our code
{\em Surya} is capable of reproducing the results of Dikpati \&
Charbonneau (1999).

\begin{figure}
\centering{\includegraphics[height=4cm,width=8cm]{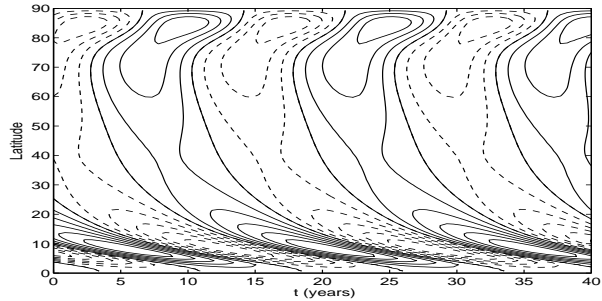}}
\caption{The theoretical butterfly diagram obtained with our code
{\em Surya} for the low-diffusivity of model of Dikpati \&
Charbonneau (1999), except that we have taken $u_0 = 20$ m
s$^{-1}$.}
\end{figure}

Comparing with the high-diffusivity solution shown on the left side of
Fig.~4, we find that the poloidal field in this low-diffusivity solution
is not able to diffuse to the tachocline from the surface,
but is advected there by the
meridional circulation.  Taking the diffusion time to be $L^2/\eta$
where $L$ is the depth of the convection zone, the diffusion time in
the low-diffusivity model is larger than 250 years, but is of order 5 years
in the high-diffusivity {\em Standard1} model.
In the low-diffusivity solution, when the poloidal field produced in a cycle
is being brought to the tachocline, the poloidal field produced in the earlier
cycles are still present at the bottom of the convection zone, which is
not the case in the high-diffusivity solution where poloidal fields
produced in the earlier cycles decay away more quickly. In the low-diffusivity
case, we expect the polar field to have an effect on the sunspot production only
when it is advected to the tachocline after a time lag.
We now carry on the same numerical experiment on this
low-diffusivity model as we did on our high-diffusivity model, by stopping
the code at every minimum and changing $A$ above $0.8\Rs$ by multiplying it by
a random number as we had done to produce Figs.~5--6.  The result
for the Dikpati--Charbonneau model (with $u_0$ taken as
20 m s$^{-1}$) is shown
in Fig.~8, with the two panels plotting the strengths of the cycle
$n+1$ and $n+3$ respectively against the value of $\gamma$ at
the end of the cycle $n$.
As expected, there is no correlation in this case between the
polar field at the minimum and the strength of the next maximum occurring
only a few years later, since the
polar field would take a longer time to be advected to the region in the
tachocline where sunspots are produced.  We, however, find a very weak
correlation between the polar field and the strength of the third next
cycle. It may be pointed out that
Charbonneau \& Dikpati (2000) carried out a study by introducing stochastic
fluctuations in the model of Dikpati \& Charbonneau (1999).  In order to
generate sunspot number plots, Charbonneau \& Dikpati (2000) took the
magnetic energy in the tachocline at a latitude of $15^{\circ}$ to be
indicative of the sunspot number.  To estimate the strengths of maxima
for producing Fig.~8, we have followed the same procedure.  Fig.~9 of
Charbonneau \& Dikpati (2000) presented correlations
of the polar field with the next few maxima on introducing stochastic
fluctuations.  They also found that the polar field was not correlated
with the next maximum and the strongest correlation was obtained
with the third maximum.

\begin{figure}
\begin{center}
\includegraphics[width=7cm]{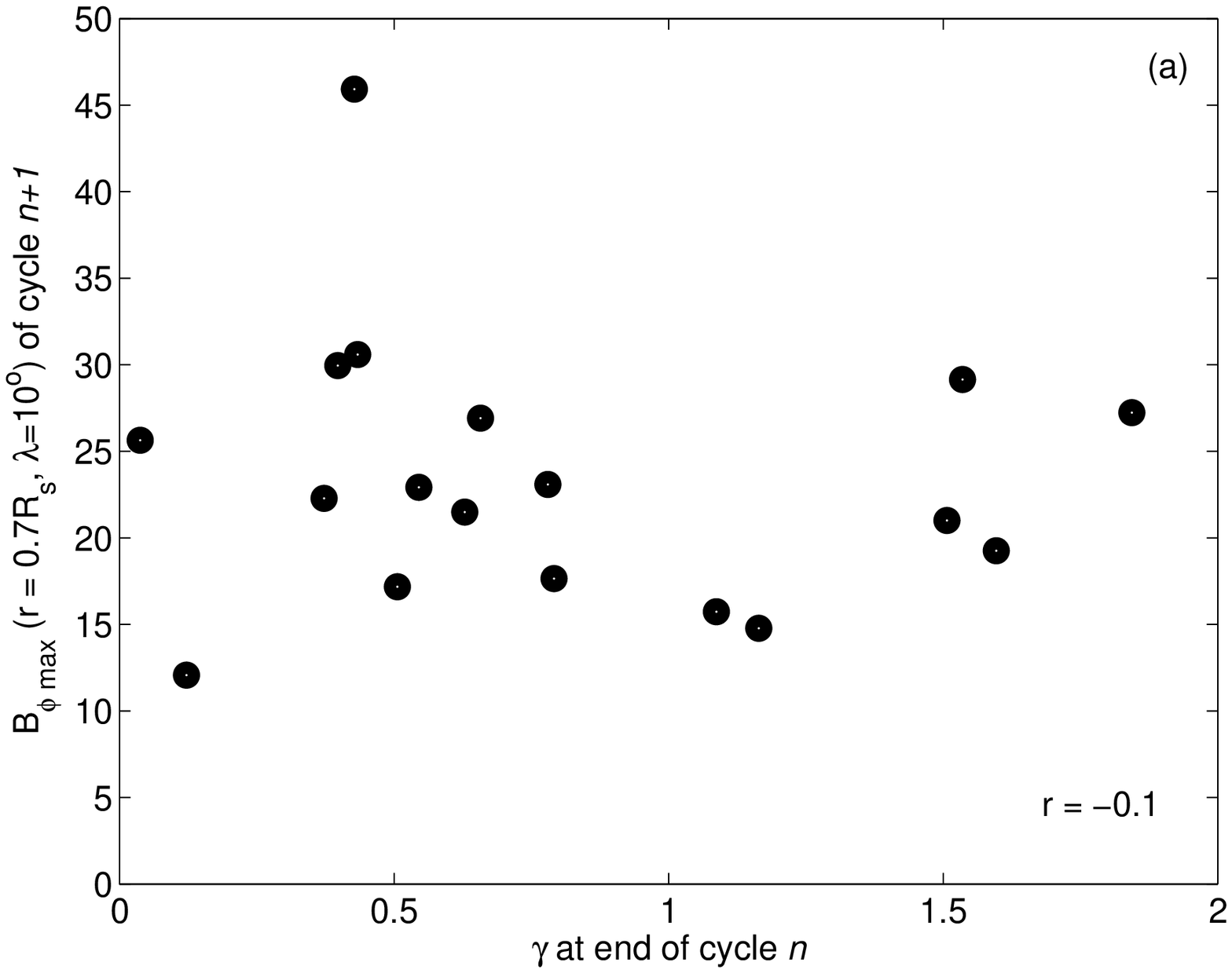}
\includegraphics[width=7cm]{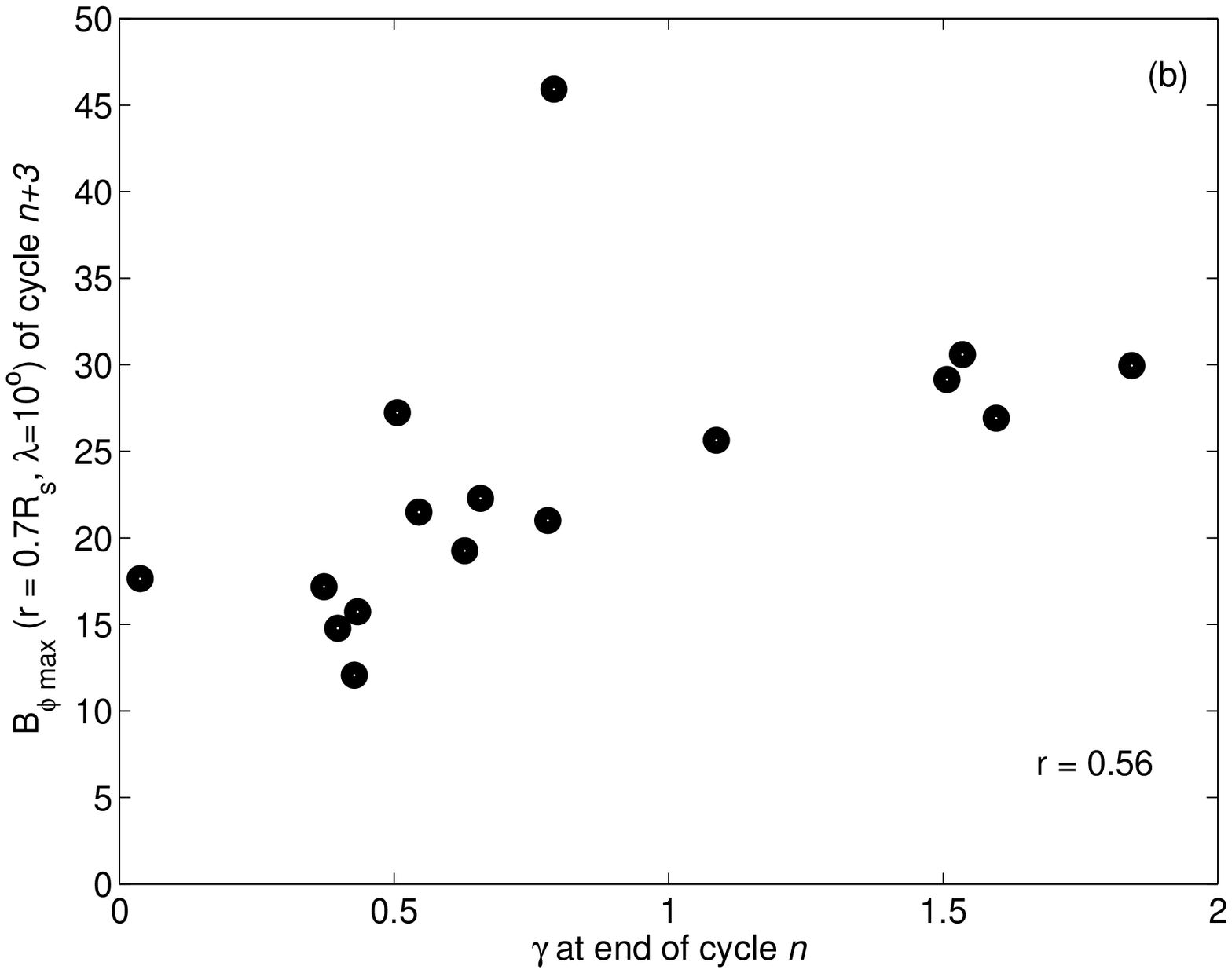}
\caption{The strength of the maximum of (a) cycle $n+1$ and (b) cycle
$n+3$ plotted against the randomly chosen value of $\gamma$ at the
end of cycle $n$. These plots are obtained from the low-diffusivity
model.}
\end{center}
\end{figure}

We now carry on the same numerical experiment with
this low-diffusivity  model as what had been done by CCJ to produce
their Fig.~2.  At a minimum, we change $A$ above $0.8$ by multiplying
it by 2.0 and 0.5 respectively.  Then we run the code without further
interruptions.  Fig.~9 shows the sunspot number plot (i.e.\ the plot
of magnetic energy in the tachocline at latitude $15^{\circ})$. We
find that a change in the polar field during the minimum has no effect
on the next maximum, has a small effect on the maximum after that
and a much bigger effect on the third maximum.  This is consistent
with the result of Charbonneau \& Dikpati (2000, Fig.~9) that
the polar field had the strongest correlation with the third
maximum. Apart from the important point
that the polar field at a minimum does not seem to have an effect
on the next maximum in a low-diffusivity model,
even the effect on the next two maxima is somewhat
smaller than the effect on the next maximum that we see in Fig.~2 of
CCJ.  It may be noted that Charbonneau \& Dikpati (2000) had to
introduce fluctuations as large as 200\% in their poloidal field
source term in order to get any noticeable effect.  It thus seems
that low-diffusivity models not only introduce an unrealistically
large time delay inconsistent with observations, they also require
very large fluctuations in the polar field to produce appreciable
effects on the dynamo.

\begin{figure}
\centering{\includegraphics[width=7cm]{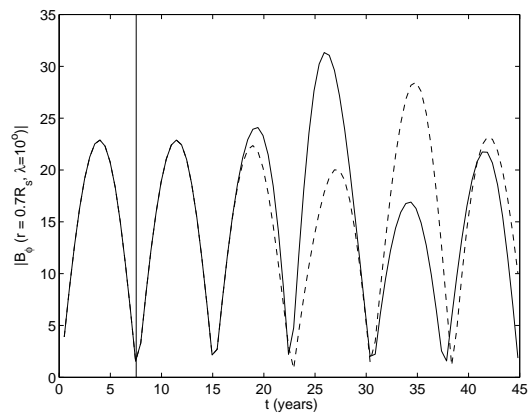}} \caption{Sunspot
number plots by increasing (solid line) and decreasing (dashed line)
the poloidal field by 50\% above $0.8\Rs$ at a solar minimum
(indicated by the vertical line).  These plots are based on the
low-diffusivity model.}
\end{figure}

The solar cycle prediction work of Dikpati \& Gilman (2006) is
based on the `calibrated' flux transport dynamo model presented
by Dikpati et al.\ (2004).  It was our intention to do some tests
on this model.  So we tried to reproduce this model with our code
{\em Surya}. We found some obvious typographical errors
in the specification of the parameters as given in Dikpati et al.\ (2004),
which were corrected (in consultation with Dr.\ Dikpati).
These typographical errors have now been listed
by Dikpati \& Gilman (2007).
Apart from these typographical corrections, we put in our code
the exactly same expressions of meridional circulation, turbulent
diffusivity and $\alpha$-coefficient as used by Dikpati et al.\ (2004)
with the same values of different parameters. We also ran the code
in a full sphere rather than in a hemisphere as done in the case
of the other calculations in this section, to make sure that all
the conditions (after correcting the typographical errors)
were identical with the conditions used by Dikpati et al.\ (2004)
to produce their `calibrated' solution. We found a decaying solution,
in contrast to the oscillatory solution which Dikpati et al.\ (2004)
reported. Although our code reproduces the results of Dikpati \& Charbonneau
(1999), the results reported in Dikpati et al.\ (2004) are not
reproduced.  That is the reason why we had to use the model of Dikpati
\& Charbonneau (1999) when we wanted to do some tests on a low-diffusivity
model.

It may be noted that Nandy \& Choudhuri (2002) argued that
a meridional circulation penetrating slightly below the tachocline
is needed to produce solar-like butterfly diagrams and we are using
such a circulation.  Dikpati \& Charbonneau (1999) and Charbonneau \&
Dikpati (2000) also used such penetrating circulation (see Fig.~2 in
Choudhuri et al.\ 2005).  Dikpati et al.\ (2004) are the only authors
to claim that they can get solar-like butterfly diagrams, with sunspots
confined to low latitudes, even with a non-penetrating meridional
circulation.  This result has so far not been reproduced by any other
group.  Other groups using non-penetrating circulation have
always obtained butterfly diagrams extending to fairly high
latitudes (Bonanno et al.\ 2002; Guerrero \& de Gouveia Dal Pino 2007).
It remains to be seen whether any other dynamo
group is able to reproduce the model of Dikpati et al.\ (2004),
which we cannot reproduce with our dynamo code and
on which the predictions of Dikpati \& Gilman (2006) are based.

\section{Arguments in favour of high diffusivity of the poloidal field}

From the results presented in the previous section, it should be clear
that the diffusivity in the convection zone has to be high if the polar
field at the minimum has to be correlated with the strength of the
next maximum as seen in the observational data.  This is a very
compelling argument that the turbulent diffusivity of the convection
zone is probably high like $2 \times 10^{12}$ cm$^2$ s$^{-1}$
as taken in our model (Chatterjee et al.\ 2004;
CCJ) and not low like $5 \times 10^{10}$ cm$^2$ s$^{-1}$
as taken in the models of Dikpati and her collaborators
(Dikpati \& Charbonneau 1999; Charbonneau \& Dikpati 2000;
Dikpati et al.\ 2004; Dikpati \& Gilman
2006).  We now list several other arguments in support of a high
diffusivity.

(i) Even if we assume the turbulent velocities within the SCZ to
have rather low values like $v \approx 10$ m s$^{-1}$ and the convection
cells to be have rather small sizes like $l \approx 30,000$ km,
still the turbulent diffusivity $\approx (1/3) v l$ would turn
out to be not less than $10^{12}$ cm$^2$ s$^{-1}$.  Thus simple
order-of-magnitude estimates favour the high values of diffusivity
that we use rather than the low values used by Dikpati and her co-workers.
Parker (1979, p.\ 629) used the convection zone model of Spruit (1974)
to conclude that the turbulent diffusivity should be of order
$1$--$4 \times 10^{12}$ cm$^2$ s$^{-1}$ within the convection zone.
It may be noted that convection zone dynamo models developed in the
early years of solar dynamo research gave best results when the
turbulent diffusivity was taken to be of order $10^{12}$ cm$^2$ s$^{-1}$
(K\"ohler 1973; Moffatt 1978, \S9.12).
Interface dynamos, however, required smaller diffusivities
of order $10^{10}$ cm$^2$ s$^{-1}$ to match the observed period of
the solar cycle (Choudhuri 1990).

(ii) Wang et al.\ (1989) studied the evolution of the diffuse magnetic
field on the solar surface under the joint influence of diffusion and
meridional circulation.  They concluded that theory fits observations
best if diffusivity at the solar surface is taken to be $10^{12}$ cm$^2$ s$^{-1}$
or larger,
comparable to the diffusivity within convection zone used in our
model.  While the surface value of diffusivity inferred by Wang
et al,\ (1989) may not necessarily imply that diffusivity inside
the convection zone also has to be comparable, it is still worth noting
that the value of diffusivity used by us is so comparable to what is
needed to match surface observations. This surface value of turbulent
diffusion also follows from the fact that the granules at the surface
have sizes of the order of a few hundred km, whereas convective
velocities are of the order of 1 km s$^{-1}$.

(iii) The solar magnetic field is of dipolar nature.  A high
diffusivity allows the poloidal field lines to get connected across
the equator and establish a dipolar parity.  Yoshimura et al.\
(1984) pointed out that a high diffusivity helped in establishing a
dipolar parity even in the traditional $\alpha \Omega$ dynamo models
without meridional circulation. This effect becomes more important
in flux transport dynamos (Chatterjee et al.\ 2004).  If the
diffusivity is low, then the dynamo solutions tend to be quadrupolar
and one needs some additional ad hoc assumption like an extra
$\alpha$-effect at the bottom of the convection zone to make the
solutions dipolar (Dikpati \& Gilman 2001; Bonanno et al.\ 2002;
Chatterjee et al.\ 2004).  While we observe the generation of the
poloidal field on the solar surface by the Babcock--Leighton process
(Wang et al.\ 1989), there is no strong observational evidence for
an additional source of poloidal field in the tachocline.  A high
diffusivity within the convection zone allows us to build models of
the solar dynamo with the correct parity without invoking an
$\alpha$-effect in the tachocline. Dikpati \&
Gilman (2007) admit that the tachocline $\alpha$ is a
``noise-amplifier''. On using mean field
equations, they find that transients take very long time to die out.
Within the real Sun, there are always fluctuations around the mean
and a noise-amplifier would not allow the system to relax to a
regular behaviour, especially when the diffusivity is low.

(iv) The irregularities in the solar cycle remain highly correlated in both the
hemispheres.  In other words, stronger (weaker) solar cycles tend to
be stronger (weaker) in both the hemispheres and longer (shorter) solar
cycles tend to be longer (shorter) in both the hemispheres.  Chatterjee
\& Choudhuri (2006) studied this problem and concluded that a high diffusivity
forces the cycles in the two hemispheres to remain locked with each other
even in the presence of asymmetries between the hemispheres.  We expect that
stochastic fluctuations without having any correlation between the two
hemispheres would lead to irregularities correlated in the two hemispheres
if the diffusivity is high, but not if the diffusivity is low.  This, however,
has to be substantiated by detailed simulations, which we are carrying out
now. The observed strong correlation of cycle irregularities between
the two hemispheres will be impossible to explain with a model having
low diffusivity.

All the above arguments taken together suggest a high value of turbulent
diffusivity within the solar convection zone.  The turbulent diffusivity seems
to have a value such that the diffusion time of the poloidal field across
the convection zone turns out to be comparable to the advection time by the
meridional circulation.  At the first sight, it may seem like a coincidence
that these two time scales seem to be so comparable.  However, the meridional
circulation is supposed to be driven by the turbulent stresses in the convection
zone.  So both of these two time scales arise from the same physics of turbulence
in the convection zone and it may not be so surprising that they are comparable.

It is usually assumed that the meridional circulation plays a very important
role in flux transport dynamos.  One may wonder if a high diffusivity will
reduce the importance of meridional circulation.  The poloidal and toroidal
fields in flux transport dynamo models are produced at the surface and
in the tachocline respectively.  These fields are clearly advected by
the meridional circulation poleward and equatorward respectively.  Since
the toroidal field is confined in a narrow layer at the bottom of the SCZ,
it is essential to assume a low diffusivity there so that advective effects
are more important there than diffusive effects
(Choudhuri et al.\ 2005).  A substantially lower diffusivity
at the base of the SCZ is assumed both by us (Nandy \& Choudhuri 2002;
Chatterjee et al.\ 2004) and by the HAO group (Dikpati
\& Charbonneau 1999; Dikpati et al.\ 2004).  If diffusive effects were
more important than the advective effects,
then the dynamo wave at the bottom of SCZ would propagate
poleward (Choudhuri et al.\ 1995), in accordance with the dynamo propagation
sign rule (see, for example, Choudhuri 1998, \S16.6).  In spite of the
assumed high diffusivity of our model within the SCZ, there is no doubt
that the meridional circulation is responsible for the equatorward propagation
of the dynamo wave at the bottom of SCZ and for the poleward advection of
the poloidal field at the surface.  In the low-diffusivity model of the HAO
group, the meridional circulation additionally advects the poloidal field
from the solar surface to the tachocline.  On the other hand, the poloidal
field in our model reaches the tachocline by diffusing from the surface
where it is generated.

If cycle 24 turns to be very strong as predicted by Dikpati \& Gilman (2006),
then that will provide a very convincing argument in favour of low diffusivity,
since a high diffusivity will not allow a strong solar cycle just after a
weak polar field in the preceding minimum.  On the other hand, a weak cycle~24
as predicted by us will make the case for high diffusivity considerably more
compelling.

\section{The connection of the theoretical model with the observational input}

In the previous two sections, we have argued that the diffusivity
of the poloidal field should be high. With such diffusivity, we expect
that the poloidal field generated at the solar surface by the
Babcock--Leighton process would diffuse towards the tachocline at all
latitudes.  Consequently, the approach followed by CCJ of updating
the theoretical model with a single number (the value of DM) at
the minimum is a drastic simplification.  We now discuss how we
can feed the observed magnetic field values at different latitudes
to update the poloidal field at the minimum.

The code {\em Surya} calculates the time evolution of
$A(r,\theta,t)$, whereas the observations provide the line-of-sight
component of ${\bf B} = \nabla \times [A(r,\theta,t) {\bf
e}_{\phi}]$ at the solar surface.  The first step is to use the
observational data to calculate $A(r=\Rs, \theta, t)$ at the solar
surface during the minimum.  The relation ${\bf B} = \nabla \times
[A(r,\theta,t) {\bf e}_{\phi}]$ implies that the magnetic field has
to be divergence-free, which requires that $\int_0^{\pi} B_r (r =
\Rs, \theta,t) \sin \theta d\theta$ integrated from one pole
to the other must be zero. If errors in magnetic
field data make $\int_0^{\pi} B_r (r = \Rs, \theta,t) \sin \theta
d\theta$ non-zero, then some special care has to be exercised when
calculating $A(r=\Rs, \theta, t)$.  We shall discuss this and then
point out how we update $A(r, \theta, t)$ underneath the solar
surface.

Following Svalgaard et al.\ (2005), we operationally define the solar polar
field as the net magnetic line-of-sight component measured through
the polarmost apertures at WSO along the central meridian on the
solar disk. We thank Todd Hoeksema for providing
line-of-sight magnetic field values averaged
over the Carrington Rotation (CR) obtained at WSO
for 30 points equally spaced in sine latitude from $-14.5/15$ to
$+14.5/15$ for each rotation from mid 1976 to 2006 (CR1642-CR2045).
The following operations have to be performed to obtain the ``real''
radial magnetic field from the WSO data (as explained to us by
Todd Hoeksema and Leif Svalgaard): (1) the raw data are
divided by the cosine of the latitude,
(2) a constant scaling factor 1.85 is multiplied to correct
the saturation effect, (3) another constant number 1.25 is
multiplied to the data of the first two years to correct the
scattered light. Thus we may obtain the radial magnetic field for
latitude -75 to +75. What we need is the value from the north pole
to the south pole. Svalgaard et al.\ (1978) pointed out that
the variation with latitude of average magnetic flux density between
the pole and the polar cap boundary obey the relation of
$\cos^8 \theta$ near the north pole and $\cos^8 (\pi - \theta)$
near the south pole. With this interpolation,
we finally obtain the values of $B_r$ between the two poles. Fig.~10 is the
contour of $B_r$ between latitudes $\pm90$ from the middle of
1976 to the end of 2006.

\begin{figure}
\centering
\includegraphics[width=90mm]{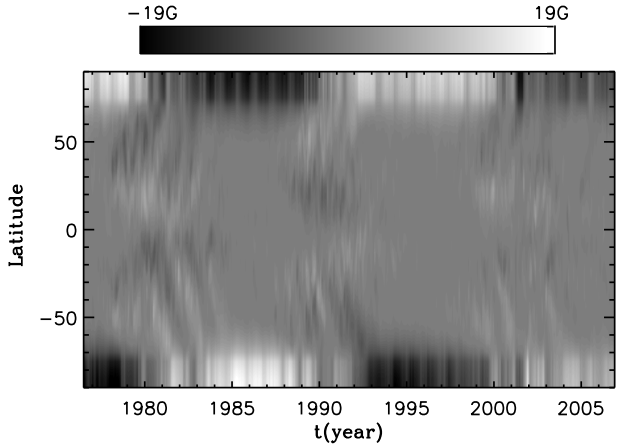}
\caption{Contours of longitude-averaged photospheric magnetic fields
from the middle of 1976 to the end of 2006 from WSO polar field
data. The values between $\pm75$ latitude are given by the
observation. The values in the polar regions are extrapolated
based on $\cos^8 \theta$ near the north pole and $\cos^8 (\pi -
\theta)$ near the south pole. During
the interval Nov. 2000 - Jul. 2002, there were some problems with
instrument sensitivity and the real fields are likely to be stronger than
that what was recorded (Schatten 2005). However, this does not pose
a problem for our theoretical modeling because we require
polar field data only during the solar minima.}
\end{figure}

The Sun's polar field reverses near the maximum of one cycle
and then begins its growth toward a new peak with opposite polarity. The
polar field becomes strong and well established about three years
before the sunspot minimum (Svalgaard et al.\ 2005). To get $B_r$ at
the minima at the ends of cycles 21, 22 and 23, we average the
raw data for the following 3-year periods just before the minima:
CR1737--CR1777
(1983:07:10--1986:06:26), CR1871--CR1911 (1993:07:03--1996:06:28) and
CR2006--CR2045 (2003:08:02--2006:07:01). For
the minimum at the end of cycle 20, however, we only have the
data since CR1642
(1976:05:27). Hence, for this case, we average the data for
8 Carrington Rotations, i.e.\ CR1642--CR1649(1976:05:27--1976:12:31).

\begin{figure}
  \begin{center}
\includegraphics[width=70mm]{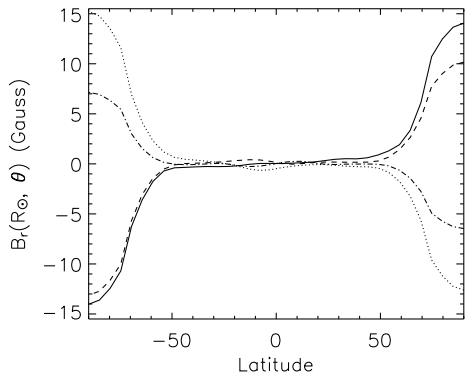}
\caption{The radial magnetic field $B_r$
for the minima at the ends of cycles 20 (solid line), 21 (dotted line), 22 (dashed
line) and 23 (dot-dashed line).  This plot is obtained from the WSO data
shown in Fig.~10. }
  \end{center}
\end{figure}

Fig.~11 shows $B_r$ as a function of latitude at the minima at
the ends of cycles 20, 21, 22 and 23.  Because of the averaging over
3-year periods, the data look reasonably smooth.  However, $B_r$
does not appear very symmetric at some of the minima.  For example,
for the minimum at the end of cycle~21, the south polar field is
clearly stronger than the north polar field.  Since the magnetic
field cannot have a monopole component, the flux must be distributed
in such a way that $\int_0^{\pi} B_r (r = \Rs, \theta,t) \sin \theta
d\theta$ turns out to be zero.  The observational data give the
following values of this quantity at the ends of the 4 successive
minima we are considering: 0.14, -0.001, 0.09 and 0.011 G. These
values give an estimate of the monopole component introduced due to
the errors in the data. Since this monopole component is not
divergence-free, it causes problems when we try to go from
$B_r(r=\Rs, \theta,t)$ to $A(r=\Rs, \theta, t)$.  To have an idea
how large the monopole components are at the various minima, keep in
mind that a uniform radial field of 1 G over the entire solar
surface would make $\int_0^{\pi} B_r (r = \Rs, \theta,t) \sin \theta
d\theta$ equal to 2 G.

To calculate $A(r=\Rs, \theta, t)$ from $B_r$, we use the relation
$$
B_r=\frac{1}{r\sin\theta}\frac{\partial}{\partial\theta}(\sin\theta
A). \eqno(4)$$
If $A$ is non-zero on any of the poles, then some terms in (1) would
become singular.  To ensure that $A$ remains zero at both the poles,
we use the following relations to obtain the surface values of $A$ in
the two hemispheres:
$$
A(R_\odot,\theta,t)\sin\theta = \left\{
\begin{array}{ll} \int_{0}^\theta B_r (\Rs, \theta', t)
\sin\theta ' d\theta' & \quad 0 < \theta< \frac{\pi}{2} \\
\int_\pi^{\theta} B_r (\Rs, \theta', t) \sin\theta' d\theta' & \quad
\frac{\pi}{2} < \theta< \pi.
\end{array} \right.\eqno(5)
$$
We take $\Rs$ as the unit of length in the above expressions giving
the numerical value of $A \sin \theta$. Only if there is no monopole
contribution, we expect $A(\Rs, \theta,t)$ calculated in the two
hemispheres to match across the equator. Otherwise, we have to
connect across the equator by the polynomial fit.

Fig.~12 shows the plots of $A \sin \theta$ as a function
of latitude at the four minima we are considering.  The plots before
smoothing and after smoothing at the equator are shown in
Figs.~12(a) and (b) respectively.
Both $B_r$ plots in Fig.~11 and
$A\sin\theta$ plots in Fig.~12 indicate that the polar field during
the minima at the ends of cycles 20 and 21 had nearly the same
strength and they were larger than that at the end of cycle 22. The
minimum at the end of cycle 23 has the weakest strength.  It may
be noted that the integrands in (5) have $B_r$ multiplied by $\sin
\theta$, which is very small in the polar regions.  Hence
uncertainties in the value of magnetic field in the polar regions do
not have much effect in the computation of $A \sin \theta$. In spite
of the errors in the data, we find in Fig.~12(a) that the jumps
in the plots at the equator are not very large.

\begin{figure}
  \begin{center}
\includegraphics[width=76mm]{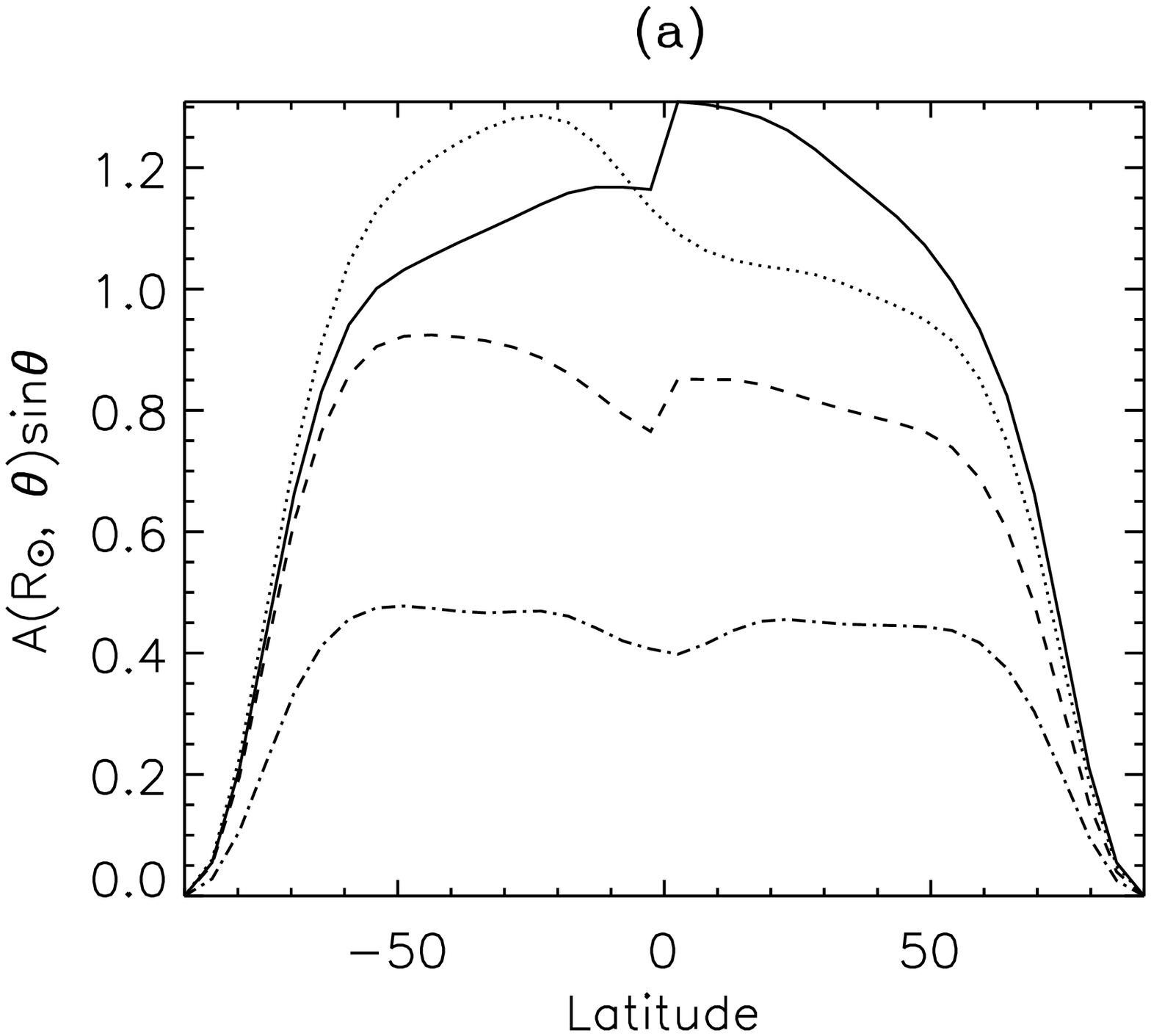}\\
\includegraphics[width=76mm]{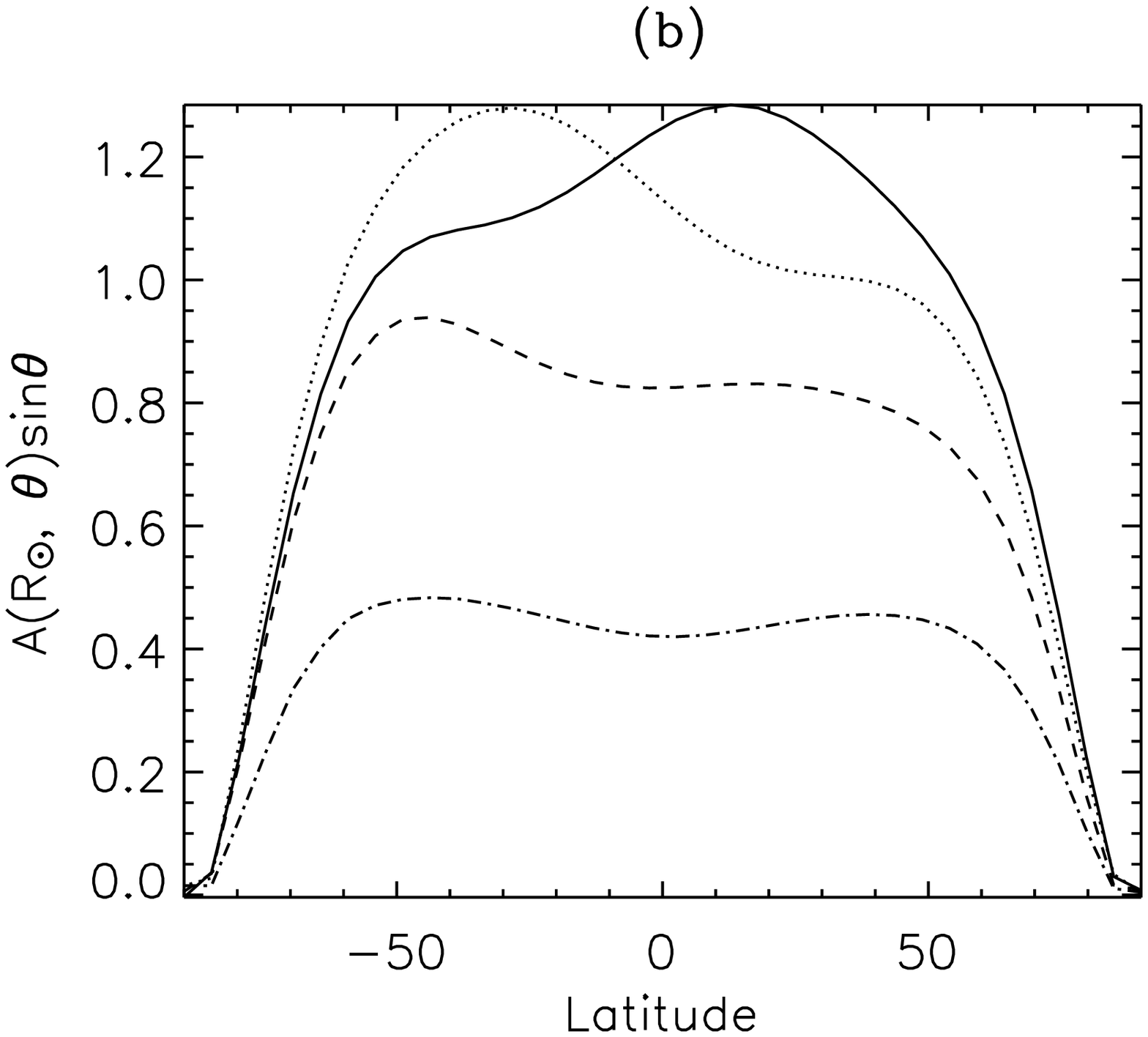}
\caption{Poloidal field $A(\Rs, \theta) \sin \theta$ inferred from
the WSO data (a) before smoothing and (b) after smoothing, for the
minima at the ends of cycles 20 (solid line), 21 (dotted line), 22
(dashed line) and 23 (dot-dashed line). It may be noted that $A(\Rs,
\theta) \sin \theta$ during the minima at the ends of cycles 21 and
cycle 23 is negative. We have plotted the absolute values.}
  \end{center}
\end{figure}

\begin{figure}
  \begin{center}
\includegraphics[width=76mm]{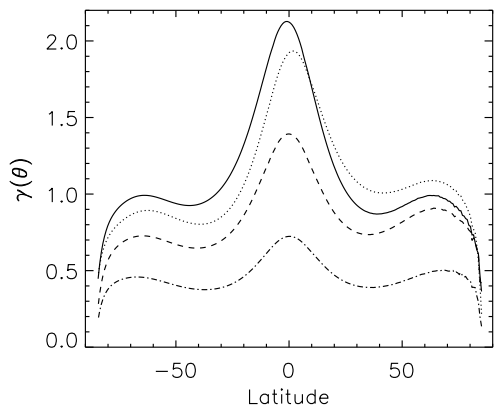}
\caption{$\gamma(\theta)$ at different latitudes for the minima at the ends
of cycles 20, 21, 22, 23. The line styles are the same as in Figs.~11 and
12. We do not change values of $A$ in the polar regions (within 5
degrees of the poles). Hence, there are no values near the poles.}
  \end{center}
\end{figure}

We now discuss how the observational data can be fed into the
theoretical model to correct for the poloidal field produced at the
end of a cycle.  We already pointed out that the poloidal field at
the minimum produced in our dynamo code corresponds to an average
cycle.  CCJ identified cycle~23 as an average cycle and took the
poloidal field at its beginning as indicative of a poloidal field of
average strength in that phase of the dynamo cycle.  From the dashed
curve in Fig.~11 giving the poloidal field at the end of
cycle~22, we find that $\overline{A \sin \theta}$ averaged over
latitude is 0.66.  Now, in our dynamo code, the only source of
nonlinearity is the magnetic buoyancy and the actual value of $A
\sin \theta$ depends on the critical magnetic field $B_c$ above
which $B$ is supposed to be buoyant.  On setting $B_c = 108$ G in {\em
Surya}, we find that $\overline{A \sin \theta}$ at the solar surface
averaged over latitude turns out to be equal to 0.66 at the time of
minimum in a regular run (taking $\Rs$ as the unit of length
when calculating $A$).
It may be noted that this value of $B_c$ can be taken as a mean
value of the toroidal field beyond which magnetic buoyancy sets
in.  Flux tube simulations suggest a value of $10^5$ G inside flux
tubes (D'Silva \& Choudhuri 1993; Fan et al.\ 1993), implying a
filling factor of order $10^{-3}$ for the flux tubes.  A somewhat
larger filling factor of order $10^{-2}$ was estimated in \S3 of
Choudhuri (2003). We put $B_c = 108$ G and run the dynamo code
from a minimum to the next minimum, when the poloidal field has to
be updated. After stopping the code at a
minimum, we check the values of $A$ at all the grid points on the
solar surface.  We denote these by $A_{\rm code} (\Rs, \theta)$.  We
have already discussed how to obtain $A$ at a minimum from the
observational data and have shown the plots in Fig.~11. Suppose
$A_{\rm code} (\Rs, \theta)$ has to be multiplied by $\gamma
(\theta)$ to make the product equal to $A$ obtained from
observational data at that latitude. Fig.~13 shows $\gamma (\theta)$
as a function of latitude for the four minima we are considering.
To avoid the numerical problem of dividing one small number by
another, we do not calculate $\gamma (\theta)$ within 5 degrees
of the poles where both $A_{\rm code} (\Rs, \theta)$ and $A$ become
very small. We take $\gamma (\theta)$ to be 1 within these regions.
We now feed the observational data in our theoretical model stopped at
the minimum in the following way.  At all grid points above
$0.8R_\odot$, we multiply $A$ by the factor $\gamma (\theta)$
appropriate for that latitude. We do not make any changes in the
values of $A$ below $0.8R_\odot$. This ensures that the poloidal
field in the upper layers, which has been created by the
Babcock--Leighton mechanism operating during the previous cycle,
gets corrected to the observed values, whereas the poloidal field at
the bottom of the convection zone, which may have been created
during the still earlier cycles, is left unchanged. Later when we
plot the poloidal field lines just after updating, we shall see
discontinuities at $0.8R_\odot$.  However, the discontinuities
in the field lines introduced by the discontinuity in $\gamma (\theta)$
at 5 degrees from the poles are completely insignificant and are
not visible in the plots of field lines.

\section{Prediction results}

We have described in the previous section how the observational
data of the poloidal field can be fed into the theoretical
dynamo model stopped at a minimum, to correct for the randomness
in the poloidal field generation process. We obtain a relaxed
solution of our {\em Standard1} model with our code {\em Surya}
and stop it at a minimum.  Identifying this as the minimum at the
end of cycle~20, we update the poloidal field by feeding
observational data appropriate for this minimum.  Then we run
the code in successive steps of one cycle, stopping at the
consecutive minima to update the poloidal field by feeding
the observational data.  The run after the minimum at the end
of cycle~23 generates the prediction for cycle~24.

\begin{figure}
  \begin{center}
\caption{(a) The contours of toroidal field and
(b) the poloidal field lines, at the solar minimum in an undisturbed
run of our {\em Standard1} model. The solid lines indicate
positive values of $B$ and $A$, whereas dotted lines
indicate negative values.}
  \end{center}
\end{figure}

Fig. 14 gives the contours of toroidal field and the poloidal fields
lines at a minimum during a regular run of {\em Surya} for the {\em
Standard1} model. Since the diffusivity is low within the tachocline
where the toroidal field is produced, we find that the toroidal
fields from the previous cycles are still present. It will be seen
that our theoretical butterfly diagrams show eruptions from the
previous cycle even after a new cycle has begun.  Our best
theoretical models for the solar cycle still suffer from this
defect.  On the other hand, the high diffusivity of the poloidal
field within the convection zone makes the poloidal fields produced
in earlier cycles decay away and we predominantly have the poloidal
field produced in the previous cycle present at the time of the
minimum.  Fig.~15 shows poloidal field lines at the four minima just
after they have been updated by feeding the observational data. It
is clear that the poloidal field tends to be asymmetric between the
hemispheres. For example, the poloidal field at the end of cycle~20
appears stronger in the northern hemisphere. This is consistent with
Fig.~11, where we see that the north polar field was stronger than
the south polar field during the minimum at the end of cycle~20. We
see discontinuities in poloidal field lines at $r=0.8 \Rs$ in
Fig.~15. These discontinuities get smoothed as our code advances the
magnetic field for a few weeks and cause no problems.  An
alternative procedure is to multiply $A$
at all depths by $\gamma (\theta)$. We have checked that this
gives very similar result for our
high-diffusivity model, where we do not have poloidal fields
created in the earlier cycles stored at the bottom of SCZ.
However, one gets considerably different
results if these different procedures are followed in a
low-diffusivity model. A very careful comparison of the
poloidal field lines in Fig.~14 with the field lines above $0.8\Rs$ in
the four plots in Fig.~15 shows that the field lines connect
across the equator more smoothly in Fig.~15 (i.e.\ the field
lines appear more bent in Fig.~14).  In other words, poloidal
field lines reconstructed from observational data suggest slightly
more diffusion between the two hemispheres compared to the
field lines from the pure theoretical model shown in Fig.~14.
This can be taken as another evidence that the magnetic diffusivity
assumed by us is not unreasonable.  If anything, the comparison
of field lines in Figs.~14 and 15 suggests that the actual
diffusivity may even be higher than what we are using in
our theoretical model.

\begin{figure}
  \begin{center}
\caption{The poloidal field lines given by constant contours of
$Ar\sin\theta$ just after correcting by the observational data
during the minima at the ends of cycles (a) 20, (b) 21, (c) 22 and
(d) 23.  The dashed lines correspond to $r = 0.8R_\odot$.}
  \end{center}
\end{figure}

\begin{figure}
  \begin{center}
\includegraphics[width=80mm]{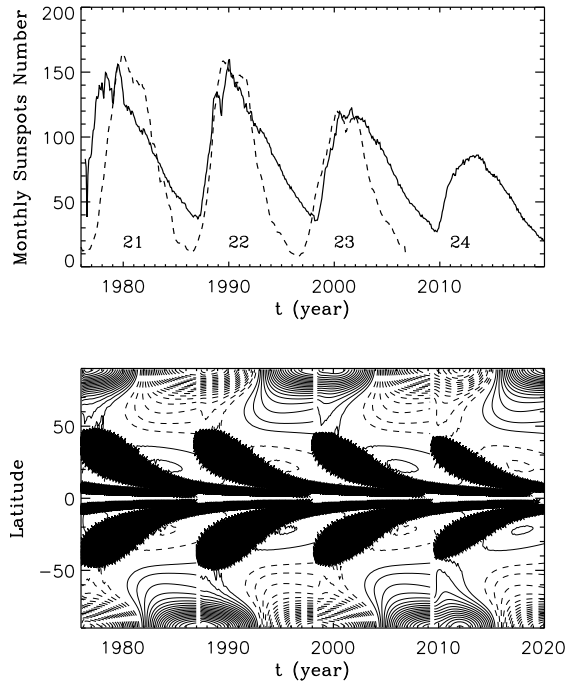}
\caption{The upper panel shows the theoretical monthly smoothed sunspot number
(solid line) superposed on the monthly smoothed sunspot numbers from
observation (dashed line). For cycles 22 and 23, the two lines match
quite well. But for cycle 21, the result from the model is slightly
weaker than the observational
result. Incomplete observational data at the end of cycle 20
might be the reason. The cycle 24 is clearly
very weak. The lower panel shows the theoretical butterfly
diagram of sunspots superimposed on the contours of radial
field for the cycles 21--24. The solid and dashed curves in
the lower panel indicate positive and negative values of $B_r$
at the surface. (Note that the solid and dashed curves got
accidentally interchanged in Fig.~3 of CCJ.)}
  \end{center}
\end{figure}

Fig.~16 now presents our results for cycles 21-24 generated by
our methodology. The top panel superposes the monthly sunspot
number generated from our model (solid line) on the observational
data (dashed line).  We may point out that the
absolute value of the theoretical sunspot number from our
numerical code does not have any particular significance, since
a finer grid would make $B$ buoyant at a larger number of grid points
and will increase the number of eruptions in our method of
treating magnetic buoyancy.  To
generate the top panel of Fig.~16, we scaled the theoretical sunspot number
suitably to make it fit the observational plot. The bottom panel
shows the butterfly diagram produced by our model. We see in the
top panel that the theoretical plot is in quite good agreement
with the observational data for cycles 22--23. The fit is not so
good for cycle~21. One possible reason for this is the incompleteness
of the poloidal field data at the end of cycle~20.  The data are
available only from the middle of that minimum and there were
some calibration problems in the first 2 years.  The other possibility
is that, after we initialize the code by feeding observational
data at a minimum, it may take about a cycle before the code
starts giving really reliable outputs. The cycle~24
comes out as the weakest cycle in a long time. The relative total
sunspots numbers for cycles 21--24 produced by the model are 3866,
3862, 3300 and 2292, respectively. Hence, the coming cycle 24
should be 30.5\% weaker than cycle 23, which makes cycle~24 a little
stronger than what it was in the earlier calculation of CCJ, who
found cycle~24 to be about 35\% weaker compared to cycle~23. We do
believe that the methodology adopted in this paper is a more
realistic, thorough and reliable methodology.  However, the
advantage of the methodology of CCJ is that this methodology,
based on using a single number like the value of DM to update
the poloidal field at the minimum, is extremely easy and
straightforward to implement in a dynamo model.  The fact that the
two methodologies give reasonably similar results suggests that
the simple method of CCJ can be used for a quick first calculation
of solar cycles, to give reasonably reliable results.

\begin{figure}
  \begin{center}
\includegraphics[width=70mm]{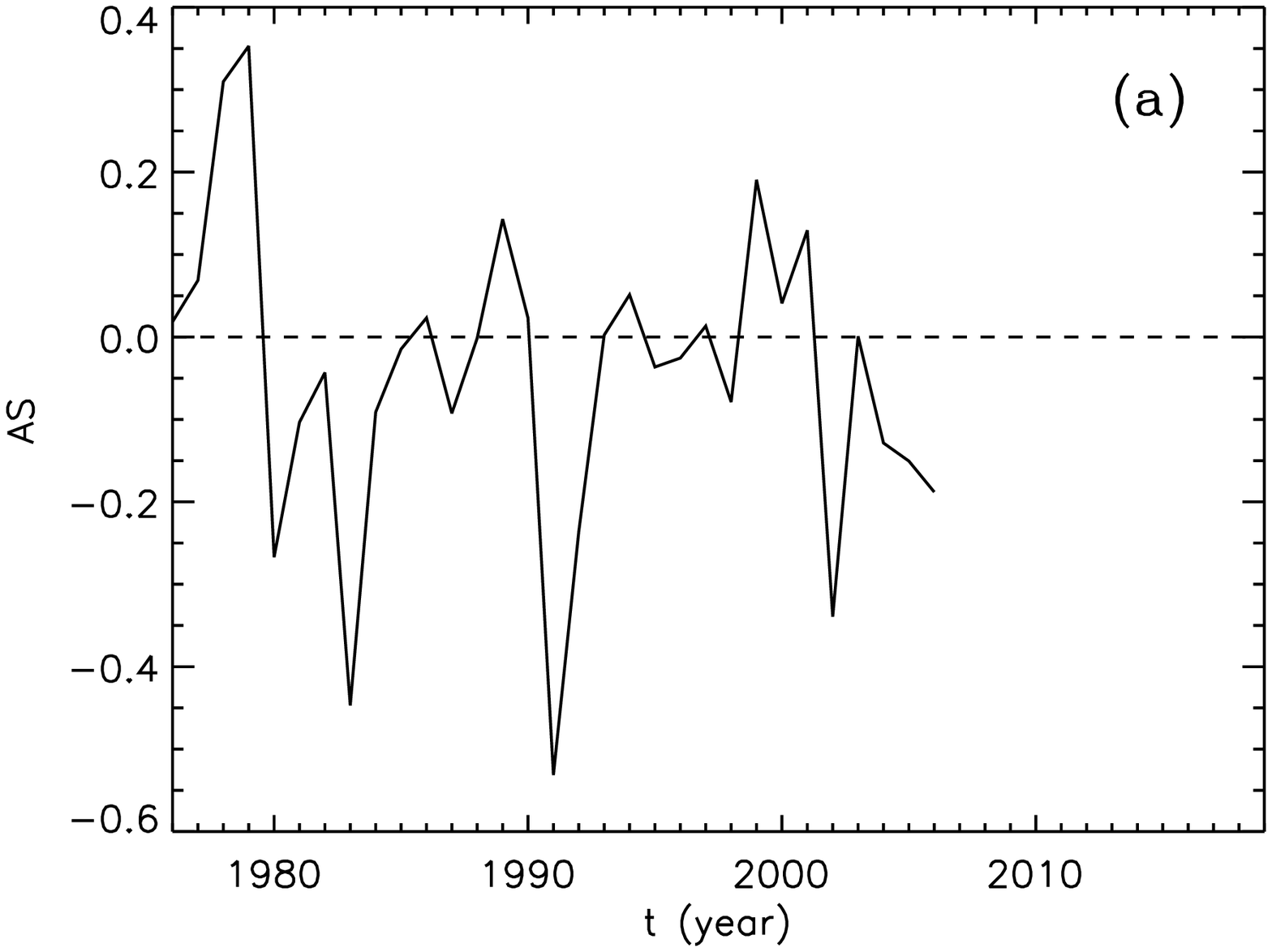}
\includegraphics[width=70mm]{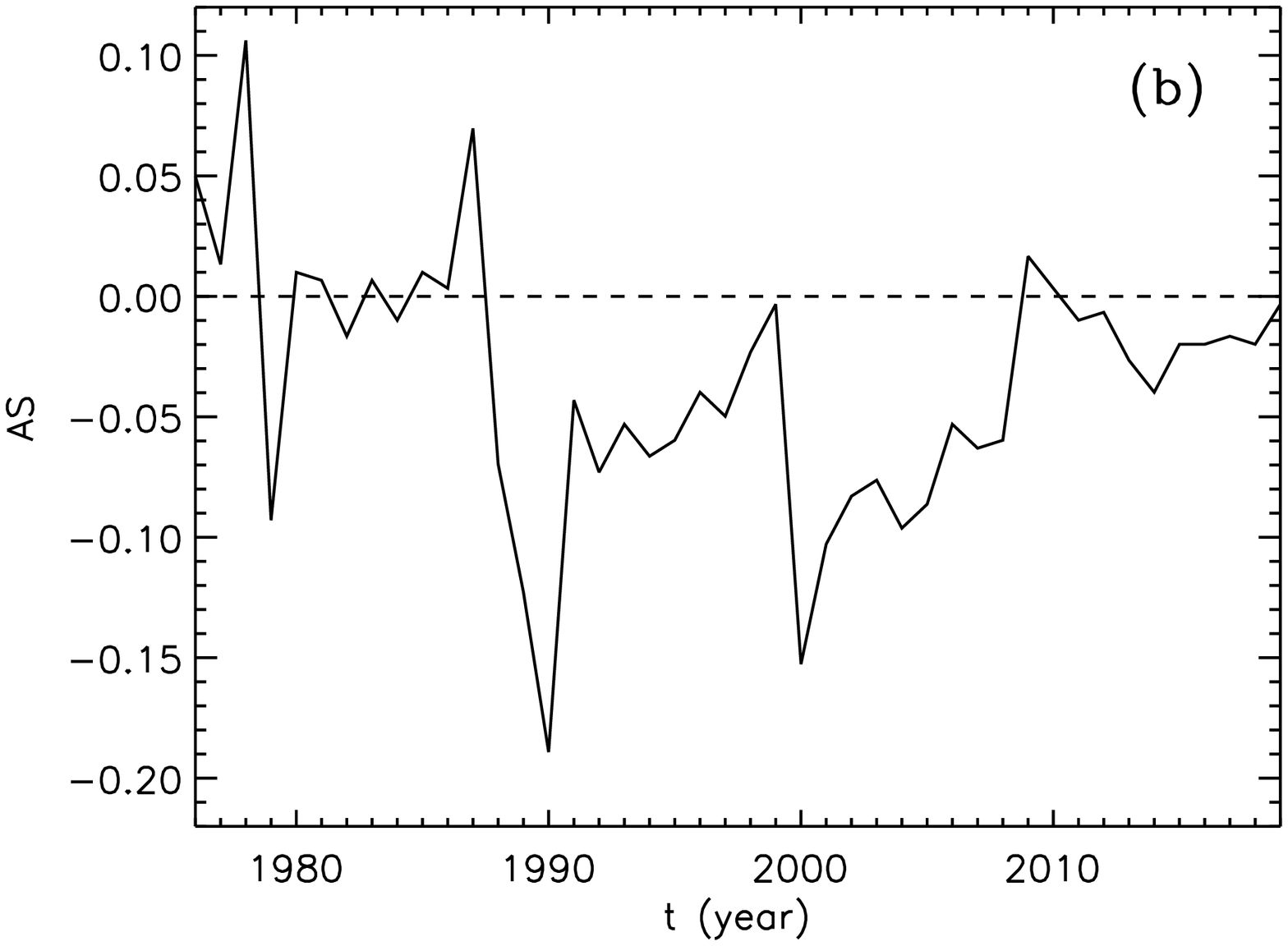}
\caption{(a): Asymmetry $AS$ of yearly sunspots area from the
observational data.  (b): Asymmetry $AS$ of yearly sunspots number from
the theoretical model.}
  \end{center}
\end{figure}

One additional advantage of our present methodology over the methodology
of CCJ is that the present methodology allows us to study the
asymmetry between the two hemispheres, which was not possible
with the methodology of CCJ. The North-South
asymmetry can be defined as
$$
AS=\frac{N-S}{[N+S]_{ave}}, \eqno(6)
$$
where $N$ and $S$ stand for the annual sunspot group numbers in
northern and southern hemisphere respectively (Li et al.\ 2002),
and $[N+S]_{ave}$ is the value of $N+S$ averaged over certain
interval, which we take here to be 1976--2006.
Different observational manifestations of solar activity, such as
major flares, sunspot numbers, sunspot area data and so on, indicate
that the solar activity can be asymmetric about the equator.
Fig.~17a shows the asymmetry $AS$ of yearly sunspot area during
1976--2006. Fig.~17b gives the asymmetry from our theoretical model.
At the first sight, it may seem that the theory does not match
observations well.  The main reason why the two plots look so
different is that the observational data is much noisier
than the theoretical result.  It may
be noted that our method of treating magnetic buoyancy introduces
some noise in the sunspot number, as can be seen in the top panel of
Fig.~16. When we used the non-local method of treating magnetic
buoyancy in our calculations with the low-diffusivity model in \S4,
the theoretical sunspot numbers turned out to be much smoother (see
Fig.~9). Also, in our theoretical model, we are including the
cumulative effect of random fluctuations by updating the poloidal
field at the minima. However, the random fluctuations in the real
solar cycle data are much larger than in the theoretical model,
which makes Fig.~17a and Fig.~17b look different.
To show that the theory matches the observations much better
on filtering out the noise, we apply the following procedure
both to the theoretical and observational data.  We calculate
the 5-year running average of $N$ and $S$ such the values of
these in the year $y$ are taken to be averages from the year
$y-2$ to $y+2$.  Then we calculate $AS$ using these running
averages.  Fig.~18 shows the theoretical plot (solid line)
superposed on the observational data (dashed line).  We now
see a much better agreement between theory and observations.
The maximum values of $|AS|$ are now comparable in the observational
and theoretical plots, which was not the case before smoothing
as can be seen in Fig.~17.

\begin{figure}
  \begin{center}
\includegraphics[width=80mm]{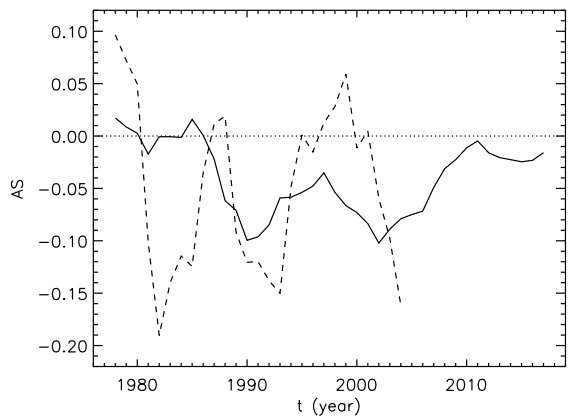}
\caption{The hemispheric asymmetries shown in Fig.~17 are now
smoothed by using 5-year running means of $A$ and $S$ before
calculating $AS$. The theoretical result (solid line) is
superposed on the observational result (dashed line).}
  \end{center}
\end{figure}

It is not difficult to understand why $AS$ in the theoretical model
tends to be negative after 1987. During the minimum
at the end of cycle~21 (i.e.\ around 1986), the south polar field
was stronger than the north polar field, as can be seen in Figs.~11 and 12.
This has clearly made the southern hemisphere more active during the next
cycle, leading to a tendency of $AS$ being
negative.  During the minima at the ends of cycles 22 and 23 also, the
south polar field has been marginally stronger than the north polar
field, although the asymmetry has not been as pronounced as at the
end of cycle 21.  Our theoretical model suggests that the tendency
of $AS$ being negative will continue in cycle~24, although this
tendency will not be as strong as it was during the period 1987--2006.
This is a very clear prediction and it will be interesting to see if
this prediction turns out to be true or not. We see in Fig.~18 that
the observational value of $AS$ tended to be negative during 1988--2005
in agreement with theoretical results.  We may also point out that
the north polar field was stronger at the end of cycle~20 (i.e.\ around 1976),
which can be seen in Figs.~11 and 12.  This led to a tendency of $AS$ being
positive during the next maximum around 1979, which is seen in
Fig.~18 both in the theoretical model and
in the observational data.

We would like to stress that the quality of the polar field data is very
important in carrying out the calculations we have reported in this Section.
We have repeated our calculation of cycles 21--24 using the NSO
data of polar magnetic field (provided by David Hathaway) in addition
to the WSO data we have been using.  We find that there is no good
agreement between the theoretical results and observational data
for the past cycles 21--23 if we use the NSO data.  We get a good
agreement only when we use the WSO data.

\section{Conclusion}

Since any theoretical investigation of a complex system like the solar dynamo
should be guided by observational data, we began by looking at the observational
data carefully.  Although systematic direct measurements of the Sun's polar
field are available only from mid-1970s, other kinds of proxies throw some
light on the polar field at earlier times.  We saw that there is reasonably
good evidence that the strength of the cycle $n+1$ is strongly correlated
with the polar field at the end of cycle $n$, giving credence to the method
of predicting solar cycles by using the polar field at the preceding minimum
as a precursor (Svalgaard et al.\ 2005, Schatten 2005).  On the other hand,
the polar field at the end of a cycle is not correlated with the strength of
the cycle. These observational facts guide us in developing our theoretical
approach.

We suggest that the lack of correlation of the polar field at the end of a
cycle with the strength of the cycle, as seen in Fig.~3, is a compelling
evidence that the generation of the poloidal field involves randomness.
Since the poloidal field is produced by the Babcock--Leighton mechanism,
the physical origin of this randomness is not difficult to understand.
The tilts of bipolar sunspots on the solar surface have a large scatter
around the mean given by Joy's law, presumably caused by the interaction
of the rising flux tubes with the convective turbulence in the uppermost
layers of SCZ (Longcope \& Choudhuri 2002).  Since the poloidal field
produced in the Babcock--Leighton process depends on the tilt, this scatter
in tilts undoubtedly would introduce a randomness.  CCJ proposed that the
theoretical dynamo model should be corrected by feeding the actual data
of poloidal field produced at the end of a cycle when we want to model
actual solar cycles. We have followed this procedure in the present work
as well.

As for the strong correlation between the polar field at a minimum
and the strength of the next cycle, the theoretical explanation depends
on the fact that the poloidal field is transported to the tachocline
and then stretched by differential rotation to produce the toroidal field
responsible for the strength of the cycle.  Since these are regular
and deterministic processes, we expect a causal link to persist.  However,
in low-diffusivity dynamo models ($\eta \approx 10^{10}$ cm$^2$ s$^{-1}$),
the polar field can be transported to the tachocline only by meridional
circulation and the advection time is too large to explain the correlation
between the polar field and the immediate next cycle.  Only in high-diffusivity
models  ($\eta \approx 10^{12}$ cm$^2$ s$^{-1}$), the poloidal field created
in the previous cycle gets advected to the pole and simultaneously
diffuses to the tachocline, leading to the observed  correlation.
We have provided several other arguments that the solar dynamo has
to be a high-diffusivity dynamo.

CCJ had used a single number (the value of DM) to feed the polar
field at the minimum into the theoretical model.  We now have developed
a methodology of feeding the detailed information of the poloidal
field at different latitudes.  While this a much more satisfactory
method than the method used by CCJ, we find that the method of CCJ also
gives results in qualitative agreement with the more detailed method.
Since the method of CCJ is much easier to implement, it can be used to
obtain a quick first result. It should be kept in mind that the prediction
of the sunspot maximum at the beginning of the cycle is possible only
because the rising phase is dominated by fairly ordered and deterministic
processes like the advection/diffusion of the poloidal field and the
generation of the toroidal field by differential rotation.  The declining
phase of the cycle, when the poloidal field is produced by the
Babcock--Leighton process, involves randomness and is not predictable.
If this view is correct, then we can think of predicting a cycle only
after the declining phase of the previous cycle is over and we know
from observations how much poloidal field has been produced.  It may
thus never be possible to make a realistic prediction of the strength
of a sunspot maximum more than 7--8 years ahead in time. Bushby \& Tobias (2007)
have given several other arguments to show the impossibility of long-term
prediction in the solar dynamo.

A high-diffusivity dynamo certainly has a shorter memory compared to
a low-diffusivity dynamo.  Additionally, when the poloidal field information
is fed into the theoretical model at the time of the minimum, that also
tries to erase the memory of the dynamo.  One important question is whether the
memory of the dynamo is restricted to be less than 11 years in our
model, since the poloidal field is updated after every 11 years.  Fig.~2
of CCJ showed that the memory of a disturbance can persist for at least
2 cycles if the dynamo is not disturbed any more after giving a single
kick.  Will the introduction of randomness after a cycle erase this
memory completely?  The answer will depend on really how random the
next kick is.  There is no doubt that the Babcock--Leighton process
introduces randomness in the generation of the poloidal field.  However, the
limited data shown in Fig.~3 is insufficient for us to conclude whether
the randomness is serious enough to erase the memory completely or
whether some weak correlation still exists between the strength of
a cycle and the poloidal field produced at its end.

Svalgaard et al.\ (2005) suggested a simple relation that the maximum
International Sunspot Number $R_{max}$ of cycle $n$ will be proportional
to the value of DM at the end of cycle $n-1$, i.e.
$$
(R_{max})_n=k(DM)_{n-1}.
\eqno(7)$$
This implies a complete loss of memory of the previous cycles.
If the randomness introduced in the Babcock--Leighton process does not
completely erase the memory of the immediately preceding cycle, then, on the basis
of Fig.~2 of CCJ showing that the memory of the polar field can persist
for a couple of cycles, we would expect a more complicated functional
relationship
$$
(R_{max})_n=f[(DM)_{n-1},(DM)_{n-2}].
\eqno(8)$$
For the last few cycles, the results obtained with our dynamo model
are roughly in agreement with what one would expect on the basis of (7).
However, if two preceding cycles have been of very unequal strengths
and (8) is the correct relation rather than (7), then it is possible
that a detailed calculation based on a dynamo model may give a
result significantly different from what one would get from (7).
The theoretical sunspot number shown in Fig.~16 also indicates that
(8) may be a more correct relation.  Since we had not fed any information
about the minimum at the end of cycle~19 into our theoretical model,
the theoretical cycle~21 does not match observations so well as the
cycles~22--23 for which data for the 2 previous minima had been fed.

One important question is whether observational data can help us
in deciding about the memory of the dynamo process.  We see in Fig.~2
that the solid circles based on actual polar field measurements lie
close to a straight line, suggesting that (7) may actually be true.
However, the open circles based on more uncertain data show a scatter.
We cannot be sure whether this scatter is due to the errors in the
data or whether there is an inherent scatter in the data.  If there
is an inherent scatter, one possible explanation is that (8) is the
correct relation and that is why we do not expect all data points
in a $(R_{max})_n$ versus $(DM)_{n-1}$ plot to lie on a straight line
or a curve.  An alternative explanation, however, is also possible.
We have been tacitly assuming that the transport of the poloidal
field to tachocline and its stretching by differential rotation there
are ordered and deterministic processes.  As argued by Choudhuri (2003),
the magnetic field has to become intermittent at some stage to account
for the existence of flux tubes. This also may introduce some
randomness, making a departure from a strict causal relationship
between $(R_{max})_n$ and $(DM)_{n-1}$.

Sometimes the so-called `even-odd rule' is forwarded as an
argument in favour of the dynamo having a memory lasting for at
least 2 cycles.  There have been 6 consecutive pairs of cycles in which the
odd cycle was stronger than the preceding even cycle.  The cycle~23
broke this `even-odd rule' after more than a century.  It is
difficult to be sure whether the `even-odd rule' really exists or
whether this apparent effect was caused by the accident of statistics.
Charbonneau et al.\ (2007) have made the provocative suggestion
that this effect
is caused by the period doubling in the dynamo,
which is a nonlinear chaotic system.
If this theoretical explanation is correct,
then the randomness introduced in the Babcock--Leighton process
should not erase the memory of the preceding cycle completely.

To sum up, there is no doubt that the Babcock--Leighton process
of poloidal field generation introduces a significant amount of
randomness in the dynamo process and this has to be corrected in
a theoretical model by using actual observational data.  Whether
this process erases all the memory of previous cycles or whether
some memory still persists is a question which cannot be settled
on the basis of currently available limited observational data.  Perhaps
we should keep our minds open and allow for the possibility that
the dynamo retains some memory even after the Babcock--Leighton
process introduces a significant element of randomness.

Our model makes a very clear prediction that cycle~24 will be a
rather weak cycle. This is a quite robust prediction even if the memory of
the dynamo persists for a few years beyond one cycle, since the polar fields at the
ends of both cycles 22 and 23 have been weak. In our high-diffusivity
dynamo model, it would be completely impossible to have a weak polar
field (like what we have now) to be
followed by a very strong cycle as predicted by Dikpati \& Gilman
(2006).   Apart from the fact our two dynamo models used diffusivities
differing by a factor of 50, the second main difference is that
Dikpati \& Gilman (2006) have taken the sunspot area data as the
source of the poloidal field.  In our view, the poloidal field
generation involves randomness and cannot be calculated
deterministically from sunspot number or area data of previous
cycles.  We now have to wait for a verdict on this debate from
the Sun-god himself in about 4--5 years' time.

\section*{Acknowledgments}

We are grateful to Dirk Callebaut, David Hathaway, Todd Hoeksema, Ken Schatten,
Leif Svalgaard and Andrey Tlatov for explaining various aspects of
observational data. We thank Axel Brandenburg, Paul Charbonneau and
Dibyendu Nandy for valuable discussions.  We also acknowledge discussions
with Mausumi Dikpati during our efforts in reproducing her low-diffusivity
model.  Suggestions from the referee, Leif Svalgaard, and the Editor helped in improving
the presentation. This work began when A.R.C.\ was a Visiting Professor at the
National Astronomical Observatories in Beijing.  He would like to thank
Jingxiu Wang, Jun Zhang and other members of the group for the very warm
hospitality. J.J.\ acknowledges financial support from National Basic Research
Program of China through grant nos. 2006CB806303, 10573025 and 10603008. P.C.\ acknowledges financial
support from C.S.I.R through grant no. 9/SPM-20/2005-EMR-I.

\clearpage

\end{document}